\documentclass[apj,numberedappendix]{emulateapj}
\slugcomment{{\sc Accepted for publication in The Astrophysical Journal:} November 9, 2010}
\usepackage{amsmath}
\usepackage{multirow}
\usepackage{natbib}
\shorttitle{ENVIRONMENT EFFECTS ON SHEAR}
\shortauthors{WONG ET AL.}

\begin{document}
\title{THE EFFECT OF ENVIRONMENT ON SHEAR IN STRONG GRAVITATIONAL LENSES\footnotemark[1]}
\footnotetext[1]{This paper includes data gathered with the 6.5 meter Magellan Telescopes located at Las Campanas Observatory, Chile.}
\author{
Kenneth C. Wong\altaffilmark{2}, 
Charles R. Keeton\altaffilmark{3}, 
Kurtis A. Williams\altaffilmark{4}, 
Ivelina G. Momcheva\altaffilmark{5}, 
and Ann I. Zabludoff\altaffilmark{2}}
\altaffiltext{2}{Steward Observatory, University of Arizona, 933 North Cherry Avenue, Tucson, AZ 85721; kwong@as.arizona.edu, azabludoff@as.arizona.edu}
\altaffiltext{3}{Department of Physics and Astronomy, Rutgers University, 136 Frelinghuysen Road, Piscataway, NJ 08854; keeton@physics.rutgers.edu}
\altaffiltext{4}{Department of Astronomy, University of Texas at Austin, 1 University Station, C1400, Austin, TX 78712}
\altaffiltext{5}{The Observatories of the Carnegie Institute of Washington, 813 Santa Barbara St., Pasadena, CA 91101}

\begin{abstract}
Using new photometric and spectroscopic data in the fields of nine strong gravitational lenses that lie in galaxy groups, we analyze the effects of both the local group environment and line-of-sight galaxies on the lens potential.  We use Monte Carlo simulations to derive the shear directly from measurements of the complex lens environment, providing the first detailed independent check of the shear obtained from lens modeling.  We account for possible tidal stripping of the group galaxies by varying the fraction of total mass apportioned between the group dark matter halo and individual group galaxies.  The environment produces an average shear of $\gamma$ = 0.08 (ranging from 0.02 to 0.17), significant enough to affect quantities derived from lens observables.  However, the direction and magnitude of the shears do not match those obtained from lens modeling in three of the six 4-image systems in our sample (B1422, RXJ1131, and WFI2033).  The source of this disagreement is not clear, implying that the assumptions inherent in both the environment and lens model approaches must be reconsidered.  If only the local group environment of the lens is included, the average shear is $\gamma$ = 0.05 (ranging from 0.01 to 0.14), indicating that line-of-sight contributions to the lens potential are not negligible.  We isolate the effects of various theoretical and observational uncertainties on our results.  Of those uncertainties, the scatter in the Faber-Jackson relation and error in the group centroid position dominate.  Future surveys of lens environments should prioritize spectroscopic sampling of both the local lens environment and objects along the line of sight, particularly those bright ($I <$ 21.5) galaxies projected within 5\arcmin~of the lens.
\end{abstract}

\keywords{gravitational lensing: strong}

\section{INTRODUCTION} \label{sec:intro}
Analyses of strong gravitational lenses have been useful in probing cosmological parameters such as $H_{0}$ \citep[e.g.][]{ref64,kee97a,sah06,ogu07}, constraining properties of the dark matter halos of galaxies \citep[e.g.][]{koo06,bar09}, and uncovering substructure in those halos \citep[e.g.][]{mao98,met01,dal02}.  These studies require accurate models of the lens potential, which can have contributions not only from the main lens galaxy, but from other objects at the lens redshift or along the line of sight.  Indeed, lens models often require environmental terms, representing a tidal shear \citep[$\gamma$;][]{kee97b} and perhaps higher-order effects \citep{koc91,kee04,fad10}, in order to yield a good fit to the observed image positions and flux ratios.  If we could measure lens environments, we would have independent and direct determinations of the shear to compare with lens models as a test of their results.

Past studies have suggested that the local lens environment can have a non-negligible effect on the lens model-derived shear of a system.  Using galaxy demographics, \citet{kee00} estimate that at least 25\% of strong lenses are in group or cluster environments that could cause strong perturbations in the lens potential.  With N-body simulations, \citet{hol03} and \citet{dal04} compute the environmental shear at the likely positions of lens galaxies, although they reach somewhat different conclusions about how strong the shear should be.  \citet{hol03} find an expected value of $\gamma \approx 0.11$, which is similar to the amount of shear needed in some lens models, whereas \citet{dal04} find a lower expected value of $\gamma \approx 0.03$.  While useful, such statistical studies do not capture the richness and possible diversity of individual lens environments.

Direct calculations of the effects of lens environments require extensive observational data.  Several studies \citep{fas02,fas06,mom06,wil06,aug07,mou07,aug08,fau09,fau10,mck10,suy10a} have used observational data to estimate the effects of environment on the potentials of strong lenses.  However, these studies did not calculate the shears induced by the environment to a level where they could draw conclusions about the most significant sources of the lens potential perturbation, nor were they always able to compare the calculated shears to the results of lens models.  Using spectroscopic and photometric data (Momcheva et al. in preparation; Williams et al. in preparation) for nine strong gravitational lenses that lie in groups, we measure the environments both around the lens and projected along the line of sight (LOS) to the lens to determine the extent to which they affect the lensing potential.  We refer to the combined local environment plus LOS perturbers as the full lens environment.  Agreement between the shears inferred from the environment and those derived from lens models would demonstrate that the environment is responsible for the previously unexplained large shears found for some lenses.  Any disagreements would point to problems in either the environment treatment or lens models.

Objects projected close to the lens can contribute to the lens potential, even if they are not physically associated with the lens.  The magnitude of the LOS perturbations decrease with both increasing separation from the lens (radial distance and redshift) and decreasing mass of the perturber \citep{mom06}.  We must characterize the effects of the LOS environment as a function of both projected separation and apparent magnitude in order to estimate the point at which LOS perturbers' contributions to the lens potential become negligible.  This will better inform the observational strategies of future surveys of lens environments by placing estimates on the size and depth of a spectroscopic survey that will sufficiently characterize the most significant contributions to the lens perturbations.

Shears derived from measurements of the lens environments may be affected by a number of systematic and random uncertainties in the theoretical formalism behind our methodology, as well as by observational errors and incompletenesses in our data.  The observational uncertainties for which we explicitly account include uncertainty in the Faber-Jackson relation, errors in the projected group centroid position and velocity dispersion, and magnitude errors in our photometry.  The theoretical uncertainties include the apportionment of mass between the group dark matter halo and the halos of individual group galaxies, the form of the density profile for the group dark matter halo, and scatter in the concentration parameter of the group halo. By quantifying the relative importance of these effects, we can determine those for which we must account in future lens surveys, as well as improve the theoretical assumptions used to calculate the lens environments.

This paper is organized as follows.  Our sample of lens systems, along with details of our spectroscopic and photometric catalog, is described in \S~\ref{sec:sample}.  The lens modeling of the 4-image lenses is in \S~\ref{sec:lens_models}.  In \S~\ref{sec:methods}, we discuss our methods for measuring the lens environments, including our handling of incompleteness in our spectroscopic catalogs and additional sources of error introduced by our theoretical assumptions and observational uncertainties.  We also describe our methodology for quantifying environment effects on the lens potential.  In \S~\ref{sec:env}, we describe our treatment of the local group environment and perturbers along the line of sight to the lens, as well as the shear due to the full lens environment, including both local group and LOS perturbers simultaneously.  We present our main results in \S~\ref{sec:results} and summarize our conclusions in \S~\ref{sec:conclusions}.  In the appendices, we present the details of the shear arising from assumptions about the group and individual galaxy halo mass distributions (Appendix~\ref{app:shear_formalism}), our calculation of the truncation radii of group galaxies (Appendix~\ref{app:trunc_rad}), and the effects of radial and luminosity cuts on line-of-sight shear (Appendix~\ref{app:data_cuts}).  Throughout this paper, we assume a flat cosmology with $\Omega_m$ = 0.274, $\Omega_b$ = 0.045, $\Omega_\Lambda$ = 0.726, and $H_{0}$ = 71 km s$^{-1}$ Mpc$^{-1}$.

\section{THE SAMPLE} \label{sec:sample}
Our sample consists of nine strong gravitational lenses chosen from the full sample of 26 lenses analyzed by Momcheva et al. (in preparation).  These nine systems are those that were determined to be in galaxy groups: \textbf{Q ER 0047-2808} \citep[hereafter Q0047;][]{war96,war98}, \textbf{HE 0435-1223} \citep[hereafter HE0435;][]{wis00,wis02,mor05,ofe06}, \textbf{MG J0751+2716} \citep[hereafter MG0751;][]{leh93,leh97, ton99}, \textbf{PG 1115+080} \citep[hereafter PG1115;][]{wey80,kun97a,ton98}, \textbf{RX J1131-1231} \citep[hereafter RXJ1131;][]{slu03,mor06}, \textbf{HST J14113+5211} \citep[hereafter HST14113;][]{fis98,lub00}, \textbf{B1422+231} \citep[hereafter B1422;][]{pat92,imp96,kun97b,kin99}, \textbf{MG J1654+1346} \citep[hereafter MG1654;][]{lan88,lan89,koc00}, and \textbf{WFI J2033-4723} \citep[hereafter WFI2033;][]{mor04,eig06,ofe06}.\footnote{Three additional lens systems, SBS1520+530 \citep{bur02,aug08}, B1600+434 \citep{jau97,fas98,koo98,aug07}, and B2114+022 \citep{kin99,aug01}, may also lie in galaxy groups.  However, we exclude these from our analysis.  SBS1520 has an uncertain lens redshift, which affects its membership in a potential host group.  The published velocity dispersion of the group thought to be associated with B1600 is less than 100 km s$^{-1}$ and determined from only five member galaxies (Momcheva et al. in preparation), which makes it very uncertain.  B2114 lacks a measured source redshift and is being strongly lensed by two galaxies at different redshifts \citep{cha01}, complicating the analysis of its lens potential.  Furthermore, all three of these systems are two-image lenses, which provide little if any constraint on the shear.  Therefore, these systems are lower priority than those for which we do have lens models.}

\begin{figure*}
\centering
\plotone{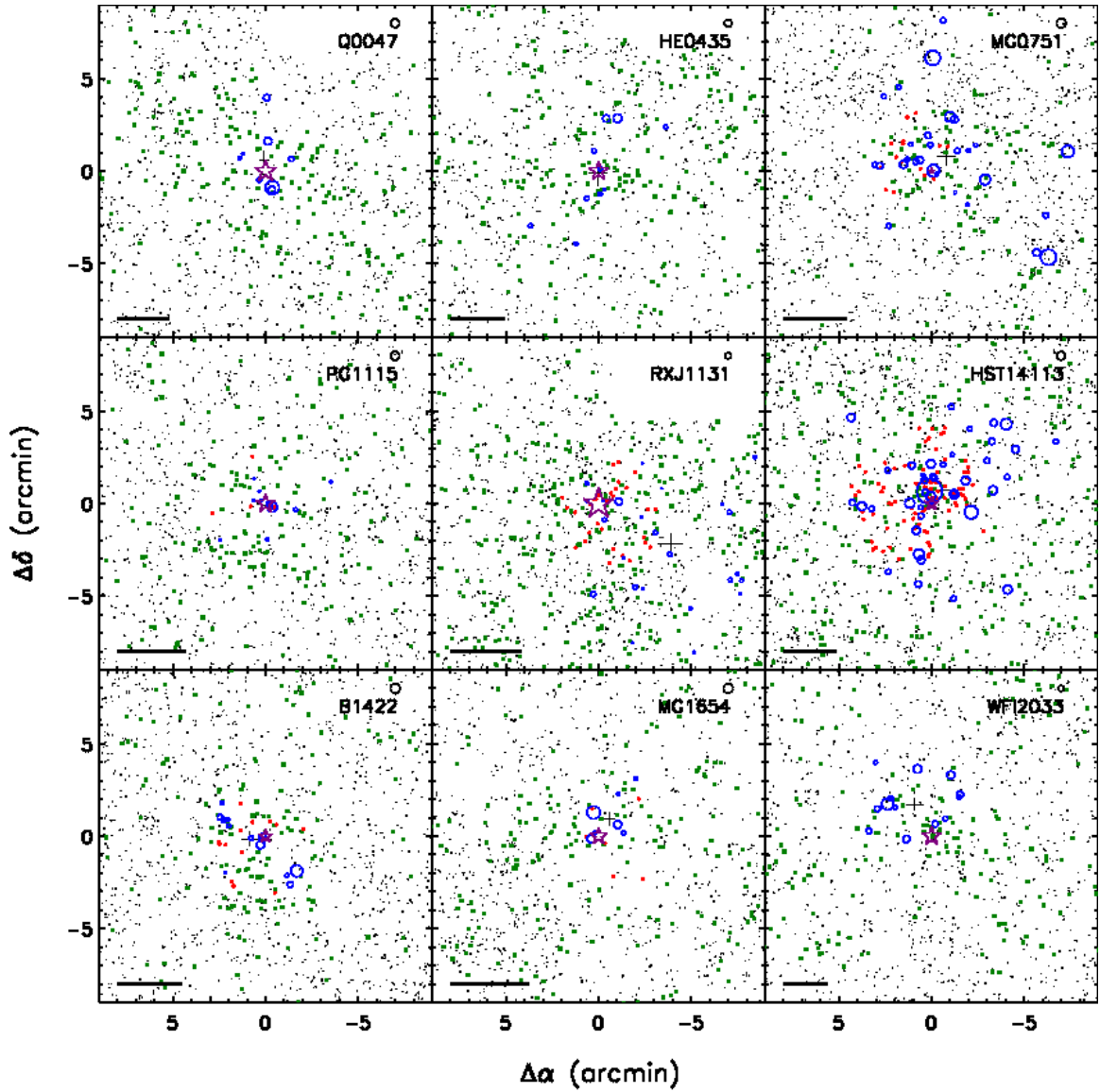}
\caption{Projected spatial distribution of galaxies in the lens fields.  North is up and East is to the left.  The fields are centered on the lens (purple star).  The black cross represents the projected group centroid and its error.  Also shown are the spectroscopically confirmed group members (open blue circles), spectroscopically confirmed line-of-sight (LOS) objects (green squares), photometric red sequence galaxies down to $I$ = 21.5 (red circles), and the remaining photometric galaxies down to $I$ = 21.5 (black dots).  The radii of the circles representing the group members are scaled by their Einstein radii, $R_{E}$.  The lens galaxy is scaled in a similar manner.  The size of the black circle in the upper right corner represents a galaxy with an internal velocity dispersion of $\sigma = 200$ km s$^{-1}$.  The area of the points representing the lens and group members scales as luminosity within a panel, but is not consistent between panels because of the $D_{LS}/D_{S}$ scaling between $\sigma$ and $R_{E}$.  The bar in the lower left corner corresponds to 1 Mpc at the lens redshift.  \label{fig:group_plots}}
\end{figure*}

\begin{figure*}
\centering
\plotone{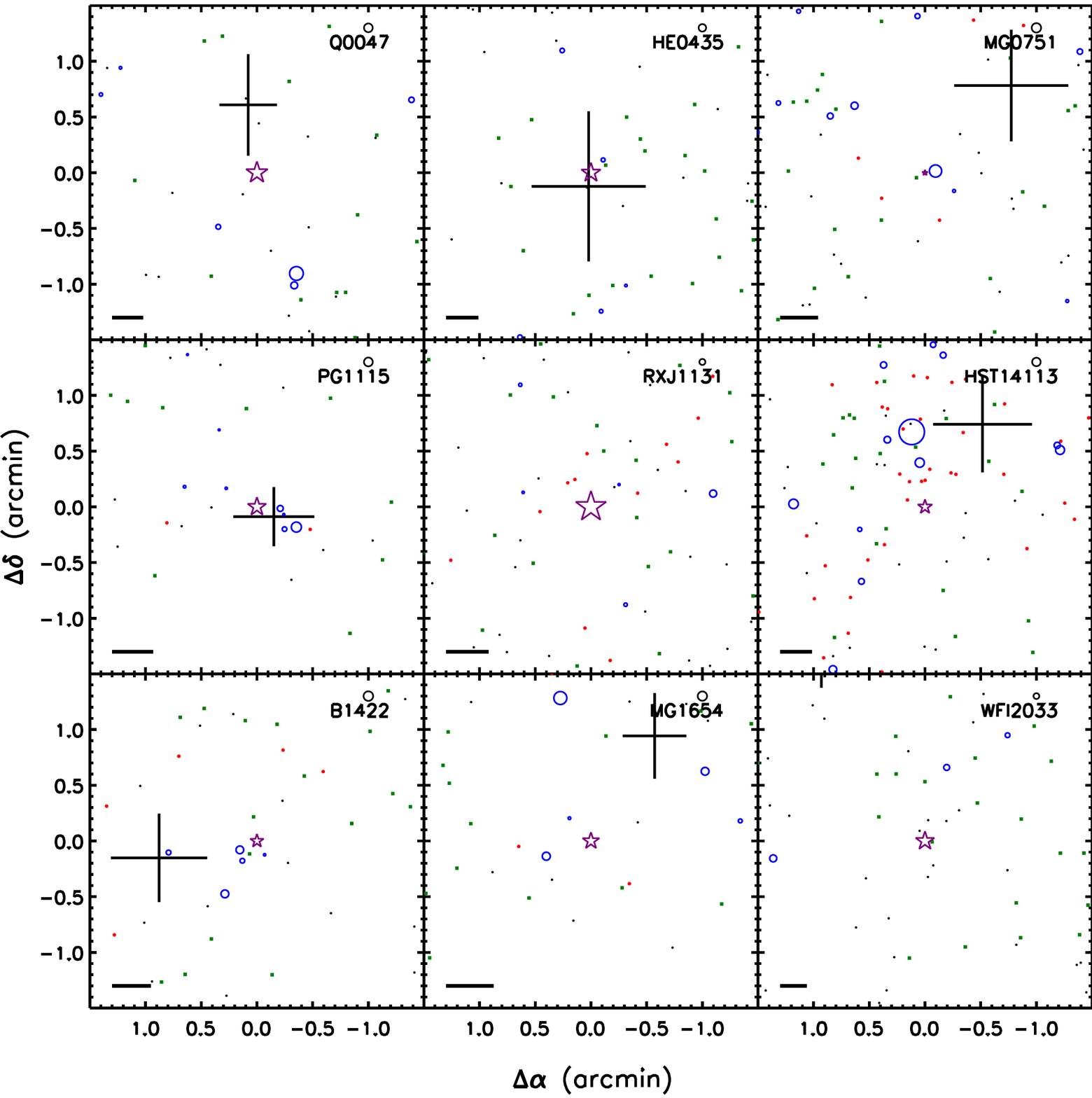}
\caption{Same as Figure~\ref{fig:group_plots}, but with a smaller field of view.  The bar in the lower left corner of each panel corresponds to 100 kpc at the lens redshift.  \label{fig:group_plots_close}}
\end{figure*}

Spectroscopic observations of four of these lenses and their environments are detailed in \citet{mom06}, with further spectroscopic data on all nine systems in Momcheva et al. (in preparation).  We also have two-band photometric catalogs of the lens fields from \citet{wil06} and Williams et al. (in preparation).  Figures~\ref{fig:group_plots} and~\ref{fig:group_plots_close} show the projected spatial distribution of objects in our catalogs.  The lens properties are in Table~\ref{tab:lenses}, and the host group properties are in Table~\ref{tab:groups}.

\begin{table*}
\begin{center}
\caption{Gravitational Lens Properties\label{tab:lenses}}
\begin{ruledtabular}
\begin{tabular}{lcccccc}
Lens &
$\alpha$ (J2000) &
$\delta$ (J2000) &
\# images &
$z_{lens}$ &
$z_{source}$ &
$R_{E} (\arcsec)$\tablenotemark{a}
\\
\tableline
Q0047 &
12.425 &
-27.874 &
ring &
0.484 &
3.60 &
1.35
\\
HE0435\tablenotemark{b} &
69.562 &
-12.287 &
4 &
0.455 &
1.69 &
1.21
\\
MG0751 &
117.923 &
27.276 &
ring &
0.350 &
3.20 &
0.35
\\
PG1115\tablenotemark{b} &
169.571 &
7.766 &
4 &
0.310 &
1.72 &
1.16
\\
RXJ1131\tablenotemark{b} &
172.965 &
-12.533 &
4 &
0.295 &
0.66 &
1.90
\\
HST14113 &
212.832 &
52.192 &
4 &
0.464 &
2.81 &
0.90
\\
B1422\tablenotemark{b} &
216.159 &
22.934 &
4 &
0.337 &
3.62 &
0.84
\\
MG1654 &
253.674 &
13.773 &
ring &
0.253 &
1.74 &
1.05
\\
WFI2033\tablenotemark{b} &
308.425 &
-47.395 &
4 &
0.661 &
1.66 &
1.16
\\
\end{tabular}
\end{ruledtabular}
\tablecomments{Lens redshifts and image separations from CASTLeS and \protect{\citet{rus03}}.}
%\tablenotetext{*}{Lens has measured time delay.}
%\tablenotetext{a}{$R_{E}$ estimated to be half image separation of lens.}
\footnotetext{$R_{E}$ estimated to be half image separation of lens.}
\footnotetext{Lens has measured time delay.}
\end{center}
\end{table*}

\begin{table*}
\caption{Host Group Properties\label{tab:groups}}
\begin{ruledtabular}
\begin{tabular}{lccccccc}
Lens &
\# members \tablenotemark{a} &
$\alpha_{cen}$ (J2000) &
$\delta_{cen}$ (J2000) &
$\alpha_{cen}$ error (\arcmin) \tablenotemark{b} &
$\delta_{cen}$ error (\arcmin) \tablenotemark{b} &
Group $\sigma_{r}$ (km $s^{-1}$) \tablenotemark{c} &
Group / RS \tablenotemark{d}
\\[1ex]
\tableline
Q0047 &
9 &
12.426 &
-27.862 &
0.26 &
0.48 &
$348^{+103}_{-79}$ &
-
\\[1ex]
HE0435 &
11 &
69.562 &
-12.290 &
0.50 &
0.66 &
$522^{+106}_{-88}$ &
-
\\[1ex]
MG0751 &
29 &
117.908 &
27.289 &
0.50 &
0.50 &
$518^{+102}_{-85}$ &
5 / 11
\\[1ex]
PG1115 &
13 &
169.568 &
7.765 &
0.35 &
0.26 &
$390^{+60}_{-52}$ &
3 / 11
\\[1ex]
RXJ1131 &
27 &
172.896 &
-12.570 &
0.64 &
0.57 &
$429^{+119}_{-93}$ &
7 / 13
\\[1ex]
HST14113 &
41 &
212.818 &
52.204 &
0.43 &
0.46 &
$656^{+66}_{-60}$ &
17 / 31
\\[1ex]
B1422 &
17 &
216.176 &
22.931 &
0.42 &
0.39 &
$421^{+99}_{-82}$ &
6 / 14
\\[1ex]
MG1654 &
8 &
253.663 &
13.791 &
0.30 &
0.38 &
$169^{+56}_{-41}$ &
2 / 11
\\[1ex]
WFI2033 &
14 &
308.449 &
-47.365 &
0.47 &
0.33 &
$498^{+84}_{-72}$ &
-
\\[1ex]
\end{tabular}
\end{ruledtabular}
\tablecomments{Tabulated values calculated using methods in Momcheva et al. (in preparation), consistent with methods described in text.}
%\tablenotetext{a}{Includes only spectroscopically confirmed members.}
%\tablenotetext{b}{Errors calculated from bootstrap resampling.}
%\tablenotetext{c}{Errors calculated from jackknife resampling and bi-weight estimators as in Momcheva et al. (in preparation).}
%\tablenotetext{d}{(\# spectroscopically confirmed group members on the red sequence) / (\# red sequence galaxies with spectroscopy).  Groups with no photometric red sequence at lens redshift are omitted.}
\footnotetext{Includes only spectroscopically confirmed members.}
\footnotetext{Errors calculated from bootstrap resampling.}
\footnotetext{Errors calculated from jackknife resampling and bi-weight estimators as in Momcheva et al. (in preparation).}
\footnotetext{(\# spectroscopically confirmed group members on the red sequence) / (\# red sequence galaxies with spectroscopy).  Groups with no photometric red sequence at lens redshift are omitted.}
\end{table*}

\begin{figure}
\centering
\plotone{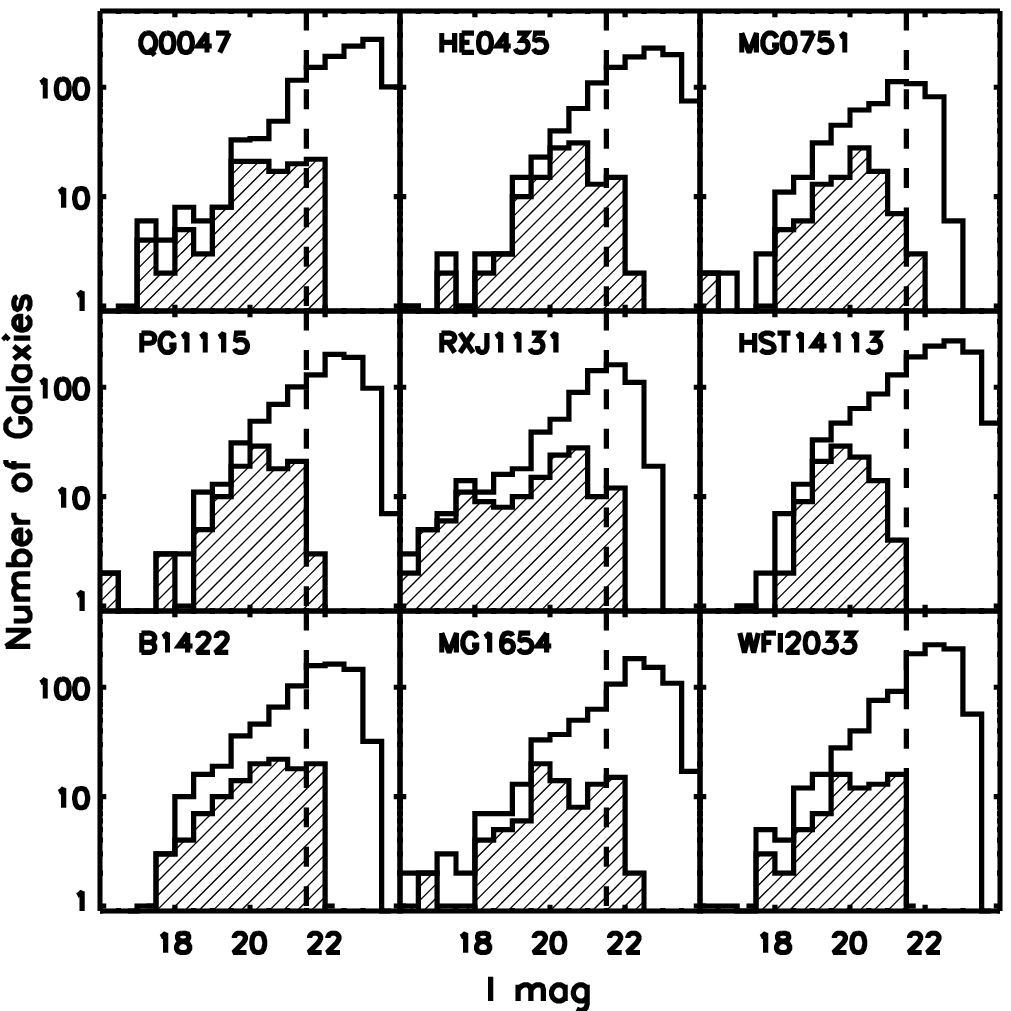}
\caption{Histogram as a function of apparent $I$-band magnitude for galaxies in the photometric sample (open) and spectroscopic subsample (hatched).  Only objects projected within 5\arcmin~of the lens are included.  Our spectroscopic completeness drops off at $I$ = 21.5 (dashed line), the limit at which we cut our sample for analysis (Appendix~\ref{app:mag_cut}). \label{fig:spec_completeness}}
\end{figure}

The spectroscopic data were taken over multiple observing runs with the Low Dispersion Survey Spectrograph-2 \citep[LDSS-2;][]{all90} and LDSS-3 on the 6.5 m Magellan 2 (Clay) telescope, as well as with the Inamori Magellan Aerial Camera and Spectrograph \citep[IMACS;][]{big03} on the 6.5 m Magellan 1 (Baade) telescope, both at Las Campanas Observatory.  Additional spectroscopic data were taken with the Hectospec multi-object spectrograph \citep{fab05} on the 6.5 m Multiple Mirror Telescope on Mt. Hopkins.  The spectroscopic target selection prioritized objects brighter than $I$ = 21.5 and within 5\arcmin~of the lens in projected separation.  We plot magnitude histograms for objects within 5\arcmin~of each lens galaxy in Figure~\ref{fig:spec_completeness}, with our spectroscopic limit of $I$ = 21.5 shown for comparison.

The photometric data are from imaging described in detail in \citet{wil06} and Williams et al. (in preparation).  Wide-field imaging is obtained for each field with the Mosaic imagers on the Kitt Peak and CTIO 4-meter telescopes in two photometric bands, Cousins $I$ and either Johnson $V$ or Cousins $R$.  Total exposure times vary in order to ensure high completeness for $I\leq 21.5$.  We obtain at least one image of each field on a photometric night.  Exposures obtained on non-photometric nights are corrected to the standard system using local stellar calibrators derived from the photometric images.  Object detection and photometric measurements are performed using SExtractor \citep{ber96}.

$I$-band magnitudes are MAG\_AUTO output from SExtractor, which are calculated in a manner similar to that of \citet{kro80}.   Total magnitudes, which we later use to calculate galaxy velocity distributions from the Faber-Jackson relation \citep{fab76}, are determined as follows.  For each field, well-isolated galaxies with no SExtractor photometric flags are selected.  We measure magnitudes in very large apertures, determine their offset from the MAG\_AUTO values, and use the mean offset to correct all objects in the catalog.  These offsets are small ($\sim 0.05$ mag) with small dispersions ($\sim 0.03$ mag).  All magnitudes used in this paper are Kron magnitudes unless indicated otherwise.

SExtractor fails to detect some galaxies, particularly those in the halos of relatively bright stars.  We manually determine positions and aperture magnitudes for those objects with $I\leq 21.5$ and within 4\arcmin~of the lens galaxy.   Total magnitudes for these galaxies are determined in a similar fashion as described above. These corrections tend to be large ($\gtrsim 0.5$ mag) and the scatter large ($\sim 0.3$ mag), as small apertures are used in order to minimize the significant shot noise from the bright star halos.

We obtain the lens galaxies' image separations from the CfA-Arizona Space Telescope Lens Survey (CASTLeS) database and from \citet{rus03}.  The HST data from CASTLeS are needed to separate the positions and fluxes of the lensed images from the lens galaxy because our photometry is not always able to resolve the individual components.

We exclude serendipitously observed stars from the spectroscopic catalog ($z < 0.01$).  We also exclude high-redshift QSOs and AGNs ($z > 1.5$).  For similar reasons, we exclude objects whose inferred $I$-band absolute magnitudes are brighter than $-25$ as galaxies this bright should be exceedingly rare \citep{bla03} given the volume of our survey.  Of all the galaxies projected within 5\arcmin~of a lens galaxy and brighter than $I$ = 21.5 across all nine fields in our spectroscopic sample, fewer than $\sim$5\% are removed by these cuts.  There is one object in the field of B1422 at $\alpha$=216.200, $\delta$=22.9475 that is spectroscopically confirmed to be a group member but does not have robust photometry because it lies under a bleed trail.  We use it in calculating the group centroid position and group velocity dispersion, but do not include it in our shear analysis.  Given its projected distance from the lens, it is unlikely to significantly affect the shear calculation.

\section{LENS MODEL SHEAR} \label{sec:lens_models}
We model the six 4-image lenses in our sample to calculate the shears needed to fit the lens data.  We consider as constraints the positions and fluxes of the lensed images, along with the position of the lens galaxy.  For HE0435, RXJ1131, HST14113, and WFI2033, we use position and relative flux data from CASTLeS, using the longest wavelength band available (typically WFPC2 F814W or NICMOS F160W) to limit the effects of microlensing and variability.  For B1422, we use radio data from \citet{pat99} to get the relative image positions and fluxes, and CASTLeS data to get the position of the lens galaxy relative to the images.  For PG1115, we take the image positions from CASTLeS, but use the mid-IR flux ratios from \citet{chi05}.  The position uncertainties are 3 mas except for the following cases: 8 mas for the RXJ1131 lens galaxy, (7,38,9,9) mas for the lensed images (A,B,C,D) in HST14113, and 9 mas for the WFI2033 lens galaxy.  We broaden the error bars on the fluxes to 10\% to account for microlensing and/or variability, except for PG1115 where we use the measurement errors reported by \citet{chi05}, since mid-IR flux ratios should not be susceptible to those systematic effects.

The models assume an ellipsoidal lens galaxy with a power law mass density profile, plus external shear.  The power law index, $\alpha$, is defined such that the enclosed mass projected within $R$ is $M(R) \propto R^\alpha$.  We do not place any explicit constraints on the lens galaxy's Einstein radius, ellipticity, and position angle, or on the amplitude and direction of the shear.\footnote{In Bayesian language, we adopt uniform priors; note that we work with quasi-Cartesian coordinates for the two components of shear, defined in Equation~\ref{eq:gamma_cs} below, and analogous coordinates for the two components of ellipticity.}  While the ellipticity and position angle have been measured for some lens galaxies, we do not want to assume that the mass necessarily traces the light.  We run Monte Carlo Markov Chains (MCMC) to sample the full range of allowed models.  For the purpose of this paper, the key model output is the distribution of shear values obtained after marginalizing over all of the lens model parameters.  We do not include any external convergence in the lens models, since that cannot be constrained due to the mass sheet degeneracy \citep{fal85,gor88,sah00}.  Including an external convergence $\kappa$ would rescale the lens models shears by $(1-\kappa)$.  We consider how this would affect our comparison of model and environment shears in \S~\ref{subsec:biases_models}.

As a fiducial case, we use all the available constraints, and we allow the lens galaxy's power law index to be a free parameter with a uniform prior in the range $0.2 \leq \alpha \leq 1.8$.  Table~\ref{tab:modshear} gives the range of recovered parameters, including the external shear, along with the $\chi^2$ goodness of fit statistics for the best models.  The $\chi^2$ values are not always comparable to the number of degrees of freedom, particularly for WFI2033, for which $\chi^{2} = 52.06$ for 5 degrees of freedom.  This suggests that there are subtleties in real lens galaxies that are not captured in our models (such as flux perturbations from substructure).  Nevertheless, we believe that the inferred shear distributions are reliable because omitting flux ratio constraints causes the shear distributions to shift by less than 2-sigma in all systems except RXJ1131 \citep[which is affected by microlensing and perhaps millilensing as well;][]{slu06,cha09,con10,dai10}.  We have also tried imposing a prior on the lens galaxy's power law index to favor isothermal profiles.  None of these systematic effects (including flux ratios in RXJ1131) changes our conclusions about whether lens model shears do or do not match the shears estimated from the observed environments, so for simplicity we report only the fiducial results.  The average shears in these six systems range from $\gamma$ = 0.06 to 0.28 for these models.

\begin{table*}
\caption{Lens Model Results \label{tab:modshear}}
\begin{ruledtabular}
\begin{tabular}{lccccccc}
Lens &
$R_{E}$ (\arcsec) &
$e$ &
PA ($^{\circ}$) &
$\alpha$ &
$\gamma_{c}$ &
$\gamma_{s}$ &
$\chi^{2}$ / DOF
\\[1ex]
\tableline
HE0435 &
$1.201^{+0.003}_{-0.002}$ &
$0.19^{+0.08}_{-0.07}$ &
$-11.7^{+2.0}_{-1.1}$ &
$0.74^{+0.33}_{-0.24}$ &
$0.064^{+0.021}_{-0.029}$ &
$-0.038^{+0.015}_{-0.011}$ &
20.49 / 5
\\[1ex]
PG1115 &
$1.138^{+0.004}_{-0.003}$ &
$0.26^{+0.04}_{-0.04}$ &
$-83.7^{+4.0}_{-4.2}$ &
$0.51^{+0.16}_{-0.12}$ &
$-0.058^{+0.015}_{-0.012}$ &
$0.149^{+0.013}_{-0.017}$ &
28.70 / 5
\\[1ex]
RXJ1131 &
$1.889^{+0.013}_{-0.018}$ &
$0.08^{+0.02}_{-0.01}$ &
$-45.0^{+8.8}_{-7.8}$ &
$1.53^{+0.14}_{-0.19}$ &
$-0.057^{+0.013}_{-0.020}$ &
$0.001^{+0.008}_{-0.012}$ &
27.71 / 5
\\[1ex]
HST14113 &
$0.839^{+0.011}_{-0.011}$ &
$0.31^{+0.11}_{-0.06}$ &
$69.6^{+11.3}_{-23.0}$ &
$1.15^{+0.23}_{-0.28}$ &
$0.272^{+0.067}_{-0.086}$ &
$0.019^{+0.027}_{-0.041}$ &
1.22 / 5
\\[1ex]
B1422 &
$0.765^{+0.023}_{-0.019}$ &
$0.36^{+0.10}_{-0.12}$ &
$-56.0^{+1.1}_{-1.8}$ &
$0.94^{+0.15}_{-0.13}$ &
$-0.047^{+0.013}_{-0.013}$ &
$-0.172^{+0.029}_{-0.029}$ &
2.22 / 5
\\[1ex]
WFI2033 &
$1.109^{+0.004}_{-0.003}$ &
$0.48^{+0.09}_{-0.09}$ &
$-87.3^{+2.4}_{-2.5}$ &
$0.74^{+0.34}_{-0.19}$ &
$0.247^{+0.018}_{-0.022}$ &
$0.136^{+0.010}_{-0.024}$ &
52.06 / 5
\\[1ex]
\end{tabular}
\end{ruledtabular}
\tablecomments{For each quantity we quote the median and 68\% confidence interval. The position angle is measured North through East. The power law index is defined such that the mass projected within radius R scales as $M(R) \propto R^{\alpha}$}
\end{table*}

\section{ENVIRONMENT ANALYSIS: AN INDEPENDENT SHEAR DETERMINATION} \label{sec:methods}
As a counterpoint to the shear inferred from lens models, we want to directly calculate the shear due to external perturbations, including both the local group environment and the galaxies along the line of sight to the lens.  Our main analysis tool is Monte Carlo simulations.  We generate many possible realizations of the local lens environment and LOS such that the shears from the different trials span the full range of shears allowed by our data.

\subsection{Overview} \label{subsec:overview}
Our goal is to use our spectroscopic and photometric data to constrain the mass distribution along the line of sight to each lens system in order to directly determine the external shear at the position of the lens.  We describe our formalism for calculating the external shear in \S~\ref{subsec:shear_determination}.  Several issues complicate this analysis.  First of all, our spectroscopy is not complete down to our spectroscopic magnitude limit, so we need to assign redshifts to those objects for which we do not have spectra.  We assign redshifts based on the redshift distribution of the spectroscopically observed galaxies of similar $I$ magnitudes in each field.  These galaxies that lack redshifts may include objects that are members of the lens host group, so we need to correct for incompleteness in the group membership as well.  We detail our methodology for handling these incompletenesses in \S~\ref{subsec:spec_incomp}.  In addition, there are a number of observational and theoretical uncertainties (\S~\ref{subsec:errors}) that we consider in our analysis.  To span a large range in parameter space including the spectroscopic incompleteness and these uncertainties, we run Monte Carlo simulations to generate multiple realizations of the lens environments.

There are two distinct components of the lens environment that we consider: the local group environment (\S~\ref{subsec:group_env}) and the LOS galaxies that are projected close to the lens but that lie at different redshifts (\S~\ref{subsec:los}).  We assume that the group halo has an NFW profile \citep[][see Appendix~\ref{app:sis_nfw}]{nav96} and that the individual galaxies are truncated singular isothermal spheres (TSIS; see Appendix~\ref{app:truncated_sis}).  The galaxy velocity dispersions used in our shear calculation are determined via the Faber-Jackson relation \citep[FJ;][]{fab76}.  We apportion mass between the group halo and the group galaxies to account for the varying degrees to which group members could be tidally stripped due to galaxy-galaxy encounters (\S~\ref{subsubsec:fhalo}).  We apply cuts on projected separation from the lens and on apparent magnitude for the LOS galaxies (see \S~\ref{subsec:los}, Appendix~\ref{app:data_cuts}).  We then calculate the shear from the full lens environment, including both the local group environment and the LOS galaxies (\S~\ref{subsec:full_env}).

For each Monte Carlo trial, we correct for spectroscopic incompleteness, apply our observational and theoretical uncertainties, and calculate the shear at the position of the lens for that particular realization.  We run 1000 trials for each lens, and the resulting shear distributions represent the possible environmental shears allowed by our data.  These distributions are then compared to the shears derived from lens modeling (\S~\ref{sec:lens_models}) to determine whether or not there is agreement between the two independent methods.

\subsection{Shear Determination} \label{subsec:shear_determination}
The full lens potential contains contributions not only from the main lens galaxy, but also from all structures along the line of sight.  For any perturber whose projected offset from the lens is larger than the Einstein radius of the lens, $R_{E}$, we can expand the lens potential in a Taylor series and quantify the effect of the perturbation by the lowest-order significant terms in the expansion, convergence ($\kappa$) and shear ($\gamma$).  $\kappa$ is a scalar quantity that represents the surface mass density of the system, but it cannot be inferred from image positions and flux ratios due to the mass-sheet degeneracy \citep{fal85,gor88,sah00}.  The shear represents tidal distortions in the lensed image and can be expressed in terms of an amplitude $\gamma$ and direction $\theta_{\gamma}$ (measured North through East).  For a spherical perturber, $\theta_{\gamma}$ represents the position angle of the perturber relative to the lens.  We can also express the shear in terms of the two components
\begin{equation} \label{eq:gamma_cs}
\gamma_{c} = \gamma \cos(2\theta_{\gamma})
\quad\mbox{and}\quad
\gamma_{s} = \gamma \sin(2\theta_{\gamma}).
\end{equation}
The total shear, $\gamma$, is their quadrature sum, $\gamma = \sqrt{\gamma_{c}^{2} + \gamma_{s}^{2}}$.  It is useful to express the shear in these terms because shears from multiple perturbers at a given redshift add linearly in ($\gamma_{c},\gamma_{s}$) space, whereas the shear amplitude $\gamma$ does not.

\newcommand\Gmat{{\bf \Gamma}}
\newcommand\Ghat{{\tilde\Gmat}}
\newcommand\vx{\mbox{\boldmath $x$}}

To determine the shear due to the full lens environment, we use the multi-plane lens equation (Schneider et al.\ 1992; Petters et al.\ 2001) to handle perturbers at different redshifts.  We discuss the full multi-plane shear formalism in a forthcoming paper (Keeton et al. in preparation) and highlight the key concepts here.  The total shear is not simply the sum of contributions from individual lens planes; there are non-linearities because each perturber acts on light rays that have already been distorted by perturbers at other redshifts.  To see this, consider the ``shear tensor'' for a perturber $i$,
\begin{equation}
  \Gmat_i = \left[\begin{array}{cc}
    \frac{\partial^2\phi_i}{\partial x_i^2} &
    \frac{\partial^2\phi_i}{\partial x_i \partial y_i} \\
    \frac{\partial^2\phi_i}{\partial x_i \partial y_i} &
    \frac{\partial^2\phi_i}{\partial y_i^2}
  \end{array}\right]
  = \left[\begin{array}{cc}
    \kappa_i + \gamma_{ci} & \gamma_{si} \\
    \gamma_{si} & \kappa_i - \gamma_{ci}
  \end{array}\right] .
\end{equation}

\newcommand\Amat{{\bf A}}
\newcommand\Imat{{\bf I}}

The shear tensors can be combined into the Jacobian matrix for the
mapping between coordinates on the sky and coordinates in the source
plane through the recursion relation
\begin{equation}
 \Amat'_j = \Imat - \sum_{i=1}^{j-1} \beta_{ij} \Gmat_i \Amat'_i \,.
\label{eq:Amat'j}
\end{equation}
where $\Amat'_1 = \Imat$ (the $2\times2$ identity matrix), and the sum
runs over all perturbers (but does not include the main lens galaxy).
Also, the weight factor $\beta_{ij}$ is a dimensionless combination of
angular diameter distances between the observer, planes $i$ and $j$,
and the source:
\begin{equation} \label{eq:beta}
 \beta_{ij} = \frac{D_{ij} D_s}{D_j D_{is}}\ .
\end{equation}
Note that $\beta_{ij}=1$ when $j$ corresponds to the source plane.
The total shear tensor is then defined by
\begin{equation}
 \Gmat_{\rm tot} = \Imat - \Amat'_s \,.
\end{equation}
If all shears are small (i.e., the components of $\Gmat_i$ are $\ll
1$), we can make a Taylor series expansion and work to first order to
approximate the total shear as
\begin{equation}
 \Gmat_{\rm tot} \approx \sum_{i} \Gmat_i \,.
\end{equation}

We account for all of the non-linear effects by constructing the full multi-plane lens equation with each perturber along the line of sight included as a complete mass component (calibrated via the methods discussed in \S~\ref{sec:env}).  This approach ensures that we properly handle both large shears from objects projected near the LOS and cumulative effects from galaxies at larger projected distances.  We use the complete multi-plane formalism to compute the shear from the full lens environment.  We also compare that with the shear obtained from the local lens environment alone to quantify the LOS effects on the shear.

The definition of the shear we have adopted for our environment analysis is chosen because it enters the lens equation in the same way as the shear inferred from lens models, and we want the two quantities to be comparable.  In principle, the lens equation can be manipulated to obtain other ``flavors'' of shear, which differ in the treatment of perturbers that lie outside the lens plane (details are given in Keeton et al. (in preparation))\footnote{For example, \citet{mom06} use the ``effective shear'', which gives reduced weight to perturbers at higher redshift offsets from the main lens.}.  Although there are formal differences among flavors of shear, we find that they lead only to small shifts in the environment shear distribution and do not change the conclusions we draw when comparing environment shears to lens model shears.  The insensitivity of our results to the shear flavor may arise because we have selected lenses that lie in group environments, which have a large component of shear from the main lens plane.

All of our shear calculations are performed using an updated version of the software developed by \citet{kee01}, which takes as inputs the relative positions of the lens and perturbing masses, as well as their Einstein radii, redshifts, and truncation radii.  The code returns $\gamma_{c}$ and $\gamma_{s}$ for the full input environment and for the local group environment alone, the latter including only objects within a small redshift $\delta z$ of the lens plane.  We set the redshifts of all lens plane galaxies, including the lens itself, to the mean redshift of all group member galaxies to eliminate redshift-space distortions due to the peculiar velocities of the group galaxies.

\subsection{Correcting for Spectroscopic Incompleteness} \label{subsec:spec_incomp}
We need to know the redshifts of the galaxies in our sample to properly compute the environment contribution to the shear.  Our spectroscopic data do not fully sample all of the objects in our photometric catalogs down to the spectroscopic limit of $I$ = 21.5 (Figure~\ref{fig:spec_completeness}), affecting our measurement of both the local lens environment and the line of sight.  There are likely group members without spectra (i.e., those that lie on the red sequence at the lens redshift), so we correct for these in considering the local environment.  We also assign redshifts to the likely LOS perturbers without spectroscopy based on the redshift distribution of our spectroscopic sample as a function of apparent total magnitude.  For each Monte Carlo trial, the local group environment includes the group dark matter halo, the spectroscopically confirmed group members, and the photometric red sequence galaxies that are assigned group membership in that trial.  The LOS perturbers include the spectroscopically confirmed non-group members in the field, the photometric red sequence galaxies not assigned group membership in that trial, and the non-red sequence galaxies without redshifts.

To correct for spectroscopic incompleteness in the host group, we consider galaxies without redshifts that lie on the photometric red sequence at the lens redshift  and assume that some are group members.  The ``red sequence'' galaxies used here and described by \citet{wil06} and Williams et al. (in preparation) are defined within a very narrow range of colors, roughly within $\pm0.05$ magnitudes, which is the typical 1$\sigma$ scatter of the red sequence \citep[e.g.][]{bow92,mci05}.  This definition is conservative in that it works to exclude interlopers at other redshifts, but it also excludes some true group members with old stellar populations that do not quite fall within this narrow color band.  All galaxies, red or blue, that are spectroscopically-confirmed group members are still accounted for in our analysis of the lens environment.  We assign each of those galaxies a probability of group membership equal to the fraction of spectroscopically observed red sequence galaxies that are confirmed group members (see Table~\ref{tab:groups}).  For each Monte Carlo realization of the environment, every red sequence galaxy has this probability of being chosen for group membership in that trial.  If a particular red sequence galaxy is not assigned group membership in a trial, we classify it as a LOS galaxy.  For different Monte Carlo trials, the red sequence galaxies assigned group membership can therefore vary in number, although every individual galaxy has the same probability of being added to the group as any other red sequence galaxies in a trial.  In three cases (HE0435, Q0047, WFI2033), no red sequence is found near the lens redshift, despite the presence of a spectroscopically confirmed group.  For these groups, we do not assign membership to any objects that are not already confirmed group members.

We assign redshifts to LOS galaxies without spectroscopic data based on the redshift distribution of our spectroscopic sample in that particular field, excluding galaxies confirmed to be in the host group of the lens.  Rather than combine the redshift distributions for all systems into an aggregate catalog, we create separate distributions to preserve the large-scale structure in redshift space in each field.  We restrict our spectroscopic sample to objects projected within 5\arcmin~of the lens and brighter than $I$ = 21.5 (little shear is contributed by objects outside these limits, see Appendix~\ref{app:data_cuts}).  We separate the galaxy redshift distributions into 8 bins of total $I$-band magnitude on a system-by-system basis.  Each bin has a width of $\sim$0.5 mag except for the first bin, which includes all objects brighter than $I \sim$ 18.  The width and locations of the bins are adjusted slightly to include at least three objects so that we can discriminate between over- and underdense regions along the line of sight.  The bin centers and widths rarely vary by more than 0.2 magnitudes in either direction.  For each bin in each system, we sort the objects by redshift and compute the cumulative probability distribution for redshift.  For each object in the photometric catalogs without spectroscopic data, we draw a random number distributed uniformly in the range [0,1] and assign it the redshift at which the cumulative distribution for the corresponding magnitude bin attains that value.

We make a slight modification when applying this redshift assignment procedure to photometric red sequence galaxies that are not selected for group membership.  For these galaxies, we again draw from a redshift distribution in bins of total $I$-band magnitude as described above.  However, because red sequence interlopers tend to be intrinsically blue galaxies at higher redshifts than the lens, we use parent distributions including only spectroscopic galaxies with $z > z_{lens}$.  This limits the number of objects in each magnitude bin, particularly for the bright ($I \sim$ 18-19) galaxies.  To maintain our condition of having at least three objects in each bin, we increase the faint-end limit of the brightest bin to $I \sim$ 19 and restrict ourselves to 6 bins instead of 8.

One caveat with this method is that the probability of selecting a redshift between two adjacent perturbers in redshift space is assumed to be uniform over that range.  This approximation is fine for well-sampled overdense regions, but does not handle voids in redshift space properly.  Overdense regions generally induce larger perturbations on the lens potential than underdense regions, so we are less concerned with voids (see \S~\ref{subsec:biases}).

\subsection{Additional Sources of Error} \label{subsec:errors}
Once all redshifts are specified, we use our observational data to build a model of the mass distribution along the line of sight.  There are uncertainties in the observational data themselves that could affect the shear, including:
\begin{itemize}
\item Slope and scatter in the Faber-Jackson relation, which we use to convert luminosities into velocity dispersions and Einstein radii.  We allow for an error of 0.20 in the slope, as well as an intrinsic scatter of $\sim$0.07 in $\log{\sigma}$ \citep{ber03a}.  These errors affect all galaxies in both the local group environment and along the line of sight.
\item Error in the group centroid position, which sets the position of the group dark matter halo.  In our shear analysis, we generate a centroid position from a bootstrap resampling of the group galaxy positions with uniform weights.  The group centroid errors are determined from the standard deviation of 1000 bootstrap trials (Table~\ref{tab:groups}).  Using luminosity-weighted centroids does not change our conclusions.
\item Error in the group velocity dispersion, which determines the total group mass.  We use a bootstrap resampling of the group galaxy redshifts and a bi-weight estimator for scale \citep{bee90} to determine the group velocity dispersion and error (Table~\ref{tab:groups}).
\item Photometric errors in the galaxy magnitudes (Williams et al. in preparation), which we use to calibrate the mass models for individual galaxies in both the local group environment and the LOS.
\end{itemize}

There are also uncertainties associated with the theoretical assumptions we must make:
\begin{itemize}
\item Choice of density profile for the group dark matter halo.  We adopt NFW profiles for our fiducial models, although we also consider a singular isothermal sphere (SIS).  Details of the difference between SIS and NFW group halos are presented in Appendix~\ref{app:sis_nfw}.
\item Scatter in the concentration parameter, $c_{vir}$, of the group dark matter halo.  We allow for a scatter of 0.14 in $\log c_{vir}$, which is approximately constant over a mass range that encompasses the virial masses of the groups in our sample \citep{bul01,wec02}.
\item The apportionment of mass between the group halo and the individual group galaxies (\S~\ref{subsubsec:fhalo}).
\end{itemize}
We build all of these uncertainties into our Monte Carlo analysis by drawing values from the appropriate distributions for each trial.  We examine the relative importance of the various uncertainties in our results in \S~\ref{subsec:results_full}.

\section{SHEAR FROM ENVIRONMENT COMPONENTS} \label{sec:env}

\subsection{Shear Contribution from Group Environment} \label{subsec:group_env}
The local group environment of each lens consists of a group dark matter halo and the individual group galaxies.  In principle, the group can contain substructure on a smaller scale than the galaxies brighter than our spectroscopic limit, but for this analysis, we will assume that its contribution is small in comparison to the overall group halo.  We only include spectroscopically-confirmed or Monte Carlo-assigned group members within a virial radius of the group centroid in each Monte Carlo trial because the virial mass of the group should not be apportioned to objects outside this radius.  We treat group galaxies outside the virial radius as LOS galaxies at the lens redshift.

\subsubsection{Shear from Group Halo} \label{subsubsec:group_halo}
In modeling the group dark matter halo, it is not clear whether virialized groups have mass density profiles that are better fit by a singular isothermal sphere (SIS) or a NFW profile \citep{nav96}.  Furthermore, if the group is not yet virialized, the form of the halo profile could be very different from either profile.  This possibility is somewhat mitigated by the fact that non-virialized groups are likely to have a larger fraction of their total mass in individual group galaxies, which may not yet have been stripped via tidal interactions with other galaxies \citep{zab98}.  We test how much of an effect the choice of an SIS or NFW profile has on the shear from the group halo and find that in the majority of cases, the effect is small, particularly when much of the group mass is apportioned to the individual galaxies rather than the overall dark matter halo.  The details of the shear calculations are presented in Appendix~\ref{app:sis_nfw}.  We assume an NFW profile throughout our analysis because the choice of group profile does not significantly affect on our results, and NFW profiles are better fits to group-size dark matter halos \citep{nav96}.

\subsubsection{Shear from Individual Group Galaxies} \label{subsubsec:gals}
We assume that each group galaxy is a SIS in the central regions, which has been shown to be an accurate model for galaxy halos \citep[e.g.][]{rus05,koo06,gav07,bol08,nip08,koo09}.  Realistic galaxy are truncated \citep[see also e.g.][]{suy10b}, so we assume that the profiles are truncated at radius $r_{t}$, as described in \S~\ref{subsubsec:fhalo} and Appendix~\ref{app:trunc_rad}.  The shear calculation for a truncated singular isothermal sphere (TSIS) depends on the internal velocity dispersion $\sigma$, from which we can determine the Einstein radius, $R_{E}$.  The details of this shear calculation are presented in Appendix~\ref{app:truncated_sis}.

For each trial, the velocity dispersions of the confirmed or assigned group galaxies are determined from their observed total $I$-band magnitudes via the Faber-Jackson relation \citep[hereafter FJ;][]{fab76}, $L \propto \sigma^{\gamma_{FJ}}$.  This procedure assumes that all group members lie on this relation (i.e. have kinematics like those of early-type galaxies).  As a result, we are effectively assigning upper limits to the masses of any late type galaxies, which tend to have lower masses than what would be calculated from this relation.  This assumption is crude, but necessary, given that we currently lack detailed morphologies for the galaxy sample.  It is not baseless, as we estimate that roughly $\sim60-80$\% of the galaxies in the host group are red and/or non-star forming (see \S~\ref{subsec:biases}).

We adopt the FJ relation given by \citet{ber03a} for the SDSS $r$-band,
\begin{equation} \label{eq:fj}
\log{\sigma} = 2.2 - \frac{0.4(M_{r}+21.15)}{3.91 + \epsilon_{slope}} + \epsilon_{int},
\end{equation}
where $\epsilon_{slope}$ and $\epsilon_{int}$ are the FJ slope error and intrinsic scatter term from \citet{ber03a}, respectively.  These terms are assumed to be Gaussian distributed with a rms equal to 0.20 and approximately 0.07 respectively for the $r$-band.  Of these two terms, the intrinsic scatter term dominates the effect on our resulting shear distributions, so future references to the effect of the FJ uncertainty on the shear reflect primarily the scatter.

We calculate the K-correction between the Cousins $I$-band and the SDSS $r$-band by synthesizing colors from an elliptical galaxy template derived using stellar population synthesis models from \citet{bru03}.  We assume a passively evolving simple stellar population with solar metallicity and a Salpeter IMF formed at $z = 3$ and evolved to the present epoch.  If the majority of our group galaxies are ellipticals, this should provide a valid approximation.  \citet{ber03a} also include a passive evolution term in the determination of $M_{r}$ that is linear in redshift.  We omit this term and account for passive evolution explicitly using the \citet{bru03} elliptical galaxy model because the range of redshifts over which we are correcting is larger than for the SDSS sample, where a linear correction is a better approximation.  For each galaxy, we evolve the model from the galaxy's redshift to $z = 0.1$, which is roughly the median redshift of the SDSS sample.  Our determination of $M_{r}$ for a particular galaxy at redshift $z_{gal}$ is 
\begin{equation} \label{eq:r_abs}
M_{r} = I_{obs} - DM(z_{gal}) + K_{I,r}(z_{gal}) + E_{I}(z_{gal},0.1),
\end{equation}
where $DM(z_{gal})$ is the distance modulus to the galaxy, $K_{I,r}$ is the K-correction term, and $E_{I}$ is the passive evolution term.  We are unable to use the Fundamental Plane \citep[e.g.][]{djo87,ber03b,cap06,rob06} because we currently do not have effective radii for the galaxies.

From this relation, we calculate the Einstein radius, $R_{E}$, for each group galaxy.  $R_{E}$ for a general SIS perturber along the line of sight is 
\begin{equation} \label{eq:r_ein_sis}
R_{E} = \frac{4 \pi \sigma^{2}}{c^{2}} \frac{D_{PS}}{D_{S}}.
\end{equation}
$R_{E}$ depends on the angular diameter distance to the source ($D_{S}$) and the angular diameter distance from the perturber to the source ($D_{PS}$).  For the group galaxies, $D_{PS}$ is equal to $D_{LS}$, the angular diameter distance from the lens redshift to the source.

The Einstein radius of each lens galaxy is approximated as half the image separation, obtained from the CASTLeS database and \citet{rus03}.  These values are within a few hundredths of an arcsecond of those determined from lens models of the 4-image lenses in our sample, so this is a good approximation.  We do not use the FJ relation for the lens galaxies themselves because our photometry is not always able to accurately deblend the galaxy from the lensed images.  We also do not use the lens galaxy fluxes from the CASTLeS webpage due to an unexplained discrepancy between those magnitudes and ours in the few systems where we can deblend our images.  In those systems, the velocity dispersions determined from applying the FJ relation to our magnitudes are consistent with both those derived from assuming $R_{E}$ as half the image separation and from direct spectroscopic measurements of the internal galaxy kinematics.  This is not true for the CASTLeS magnitudes, which generally are anomalously faint.  The properties of the lens galaxy itself do not directly enter into the calculation of the external shear, and are only used in determining the mass distribution within the group (\S~\ref{subsubsec:fhalo}).  For these reasons, we treat the lens galaxies in this separate manner.

\subsubsection{Reapportioning Mass between Group Halo and Individual Galaxies} \label{subsubsec:fhalo}
\citet{mom06} calculate the expected shear for lenses in six galaxy groups in the limits where the shear is either due to a group halo or to galaxies at the group member positions.  In our analysis, we examine the more realistic case where there is mass in both the group halo and the galaxies by redistributing the total group mass between the two components.

We define the parameter $f_{halo}$ to be the ratio of the group halo mass to the virial mass of the group (see Equation~\ref{eq:m_halo}),
\begin{equation} \label{eq:f_halo}
f_{halo} = \frac{M_{halo}}{M_{vir}}.
\end{equation}
We expect galaxy halos to be tidally stripped as the group forms and ages.  In the extreme limit $f_{halo} = 0$, all of the mass is attached to the galaxies and there is no mass in a common halo.  As we approach large $f_{halo}$, virtually all the mass has been stripped from the galaxies and merged into a common halo.  Therefore, $f_{halo}$ characterizes the extent to which the halos of group galaxies may have been tidally stripped.  Because we do not have observational constraints on $f_{halo}$, we allow it to vary as a free parameter in our Monte Carlo simulations.

For each realization of the group environment in our Monte Carlo simulations, we choose a random value of $f_{halo}$ uniformly distributed between 0 and 1 and assign masses to the group halo and the group galaxies as described above.  We distribute the total mass, $M_{vir}$, between the group halo and the confirmed or assigned group members projected within $r_{vir}$ of the group centroid.  For a particular value of $f_{halo}$, the galaxies' truncation radii, $r_{t}$, are scaled so that the density at the truncation radius is the same for all group galaxies, and the total mass of the galaxies is equal to $(1 - f_{halo})M_{vir}$.  The details are in Appendix~\ref{app:trunc_rad}.

Since the mass apportionment can vary depending on the distribution of galaxy luminosities for the group members, the values of $f_{halo}$ are not directly comparable from group to group in terms of how $f_{halo}$ affects the shear.  Similar values of $f_{halo}$ between groups do not necessarily imply that an $L^{*}$ galaxy has the same truncation radius from group to group.  For the group with the most mass per galaxy in our sample (HE0435), a value of $f_{halo}$ = 0 implies typical truncation radii of over $\sim$1 Mpc, while for the least massive group (MG1654), it implies truncation radii of $\sim$60 kpc.

For large values of $f_{halo}$, the truncation radii of the group galaxies may be smaller than their typical luminous radii, which is unphysical.  For Monte Carlo trials where this occurs, we repeat the trial with a new value of $f_{halo}$ until the truncation radii become physical.  A study by \citet{ros10} using early-type galaxies from the SDSS finds that the effective (half-light) radii of galaxies with absolute magnitudes similar to those of our galaxies are best fit by the relation $\log{(R_{eff})} = -0.257M_{r} - 5.086$.  We determine each group galaxy's absolute $r$-band magnitude, use this relation to calculate $R_{eff}$, and take $2R_{eff}$ to be the minimum truncation radius allowed.  For a de Vaucouleurs profile, this radius encloses $\sim70$\% of the total light of the galaxy.  A Monte Carlo trial is repeated with a new, smaller value of $f_{halo}$ if any galaxy in the trial is assigned a truncation radius smaller than this.  We run tests in which the minimum allowed truncation radius is $5R_{eff}$ (enclosing $\sim90$\% of the light), but this does not change our results.

\subsection{Shear Contribution from Line-of-Sight Objects} \label{subsec:los}
Objects along the line of sight to the lens also contribute to the external shear by inducing perturbations in the lens potential \citep{kee03}, even if they are not dynamically associated with the group.  Our shear calculation software can determine the contribution from a perturber at any redshift along the line of sight, making it straightforward to calculate the contributions from all LOS objects.  We apply a similar formalism to the LOS galaxies as we do for the group galaxies (\S~\ref{subsubsec:gals}).  Using Equations~\ref{eq:fj} and~\ref{eq:r_abs}, we determine their velocity dispersions (and therefore, their Einstein radii) from their observed total magnitudes and redshifts.  Our use of the FJ relation is potentially more problematic here, but we still estimate that $\sim50$\% of the LOS galaxies are red and/or non-star forming (see \S~\ref{subsec:biases}).  The truncation radii of the assumed TSIS for the LOS galaxies are approximated as $r_{200}$, where the mean density inside the volume bounded by a sphere of that radius is 200 times the mean matter density of the universe.  $r_{200}$ for an isothermal sphere at a redshift $z_{gal}$ and with a velocity dispersion $\sigma$ is 
\begin{equation} \label{eq:r_200}
r_{200} = \frac{\sqrt{2}\sigma}{10 H_{0} \sqrt{\Omega_{m} (1+z_{gal})^{3}}}.
\end{equation}

To account for all LOS perturbers, we need to correct for the spectroscopic incompleteness for objects brighter than our limit of $I$ = 21.5.  Our photometric catalog contains many galaxies brighter than this limit without spectroscopic data.  We model their redshift distribution in a physically motivated way (\S~\ref{subsec:spec_incomp}), calculate their masses and truncation radii (Equation~\ref{eq:r_200}), and determine their effects on the lens shear.

One concern is the effect of LOS objects projected far enough away from the lens that they lie outside of our sampling region.  Including objects projected far from the lens is both observationally and computationally unfeasible.  Simulations run by \citet{bra10} show that the external shear due to galaxy-galaxy lensing extrapolates to zero at a projected separation of $\sim$5\arcmin.  Observational studies \citep[e.g.][]{aug07} have suggested that the most significant perturbers are projected even closer to the lens ($\sim$15\arcsec).  Given that the shear contribution from an object decreases with increasing projected separation, we make and justify a cut at 5\arcmin~(Appendix~\ref{app:sep_cut}).

Finally, we consider the shear contribution from objects fainter than our spectroscopic limit, which is more difficult because we do not have a large spectroscopic sample at fainter magnitudes from which to build redshift distributions.  To avoid this problem, we cut our photometric sample at our spectroscopic limit of $I$ = 21.5.  We gauge the effects of fainter objects and give justification for this cut in Appendix~\ref{app:mag_cut}.  We find that a cut at $I$ = 21.5 includes the most significant perturbations and is a reasonable cutoff magnitude for our analysis.

\subsection{Shear from Full Lens Environment} \label{subsec:full_env}
The local group environment and LOS perturbers both contribute to the observed shear, although by varying amounts.  Whereas the previous sections outlined our methodology for determining their respective shear contributions, a comprehensive treatment of the shear calculation must simultaneously include all of the perturbations from the full lens environment.  Considering the LOS and local environment effects separately may be approximately correct if the effect of one component is small relative to the other.  However, we do not know \emph{a priori} whether this is the case, and if so, which component is the dominant one.  Such an analysis does not account for interactions among multiple lensing planes, so we run a full simulation with all perturbers included in order to properly handle these effects.  At a fundamental level, higher-order terms in the lens potential can be analyzed in the full analysis (Keeton et al. in preparation), but we only consider the shear term in our calculation.

In our Monte Carlo simulations of the full lens environments, we run 1000 trials for each lens.  For each trial, we assign group membership to a subset of photometric red sequence galaxies based on the fraction observed from spectroscopy (\S~\ref{subsec:spec_incomp}) and draw randomly from a distribution of $f_{halo}$ values with a uniform prior (requiring that no galaxies are assigned a truncation radius less than $2R_{eff}$).  The total group mass is then apportioned among the group halo and the individual group members (\S~\ref{subsubsec:fhalo}).  We assign random redshifts to the LOS galaxies using the procedure in \S~\ref{subsec:spec_incomp}.  We assume Gaussian errors to account for most of the uncertainties discussed in \S~\ref{subsec:errors} except when stated otherwise.  Each trial therefore represents one possible realization of the full lens environment.  By running 1000 trials, we are sampling a large region of our parameter space in order to determine the range of external shears that could be produced by the environment.

\section{RESULTS \& DISCUSSION} \label{sec:results}
\subsection{Effects of Errors on Full Lens Environment Shear} \label{subsec:results_full}
We run five sets of 1000 realizations of the full lens environments (\S~\ref{subsec:full_env}), allowing for variations due to one of our uncertainties (\S~\ref{subsec:errors}) in each set.  We also run a set of trials representing a control sample excluding these uncertainties, as well as a set of trials including all of the uncertainties simultaneously (Figure~\ref{fig:env_calc}).  The variation in the shears for the control set essentially accounts just for variations in $f_{halo}$, the selection of additional group members from the photometric red sequence, and the variations in the redshift distributions of the LOS perturbers, but for none of the additional uncertainties described in \S~\ref{subsec:errors}.

\begin{figure*}
\centering
\plotone{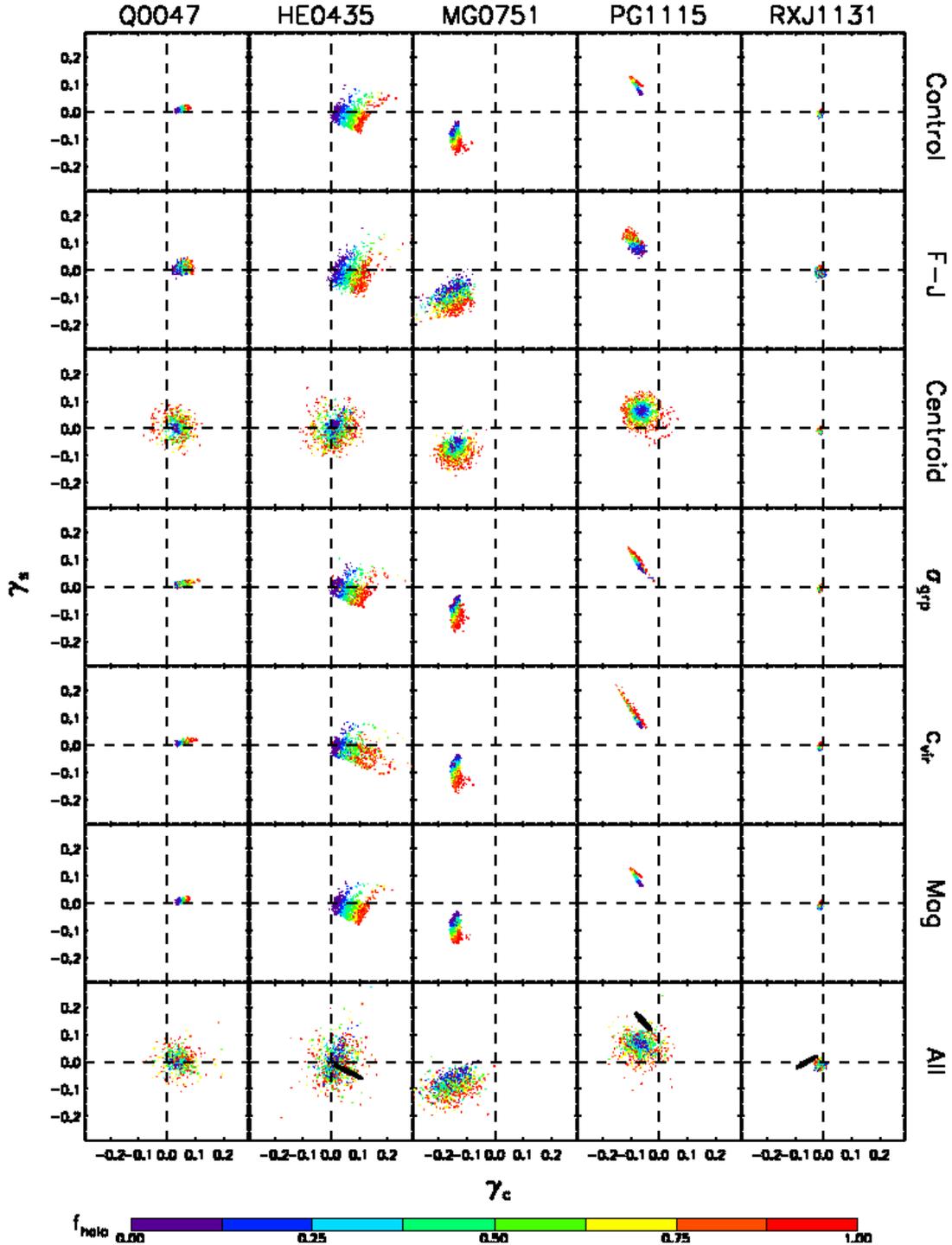}
\caption{$\gamma_{c}$ vs. $\gamma_{s}$ for the full lens environment with only systematic effects (variations in $f_{halo}$, selection of additional group member candidates from the red sequence, variations in the redshift distribution of the line-of-sight perturbers; top row), isolated observational uncertainties (2nd through 6th rows), and the systematics plus all observational uncertainties (bottom row).  The points are color coded by $f_{halo}$ as indicated by the color bar.  For the six 4-image lenses, the 1$\sigma$ and 2$\sigma$ error ellipses derived from lens modeling are shown for power-law lens galaxy models (bottom row).  The scatter in each panel is quantified in Table~\ref{tab:err}.  In three of the six 4-image lenses (HE0435, PG1115, HST14113), our environmental shears match the model shears, but in the others (RXJ1131, B1422, WFI2033), the environmental and model shears are inconsistent at the $> 95$\% level (\S~\ref{subsec:env_mod_comp}).  \label{fig:env_calc}}
\end{figure*}

\addtocounter{figure}{-1}
\begin{figure*}
\centering
\plotone{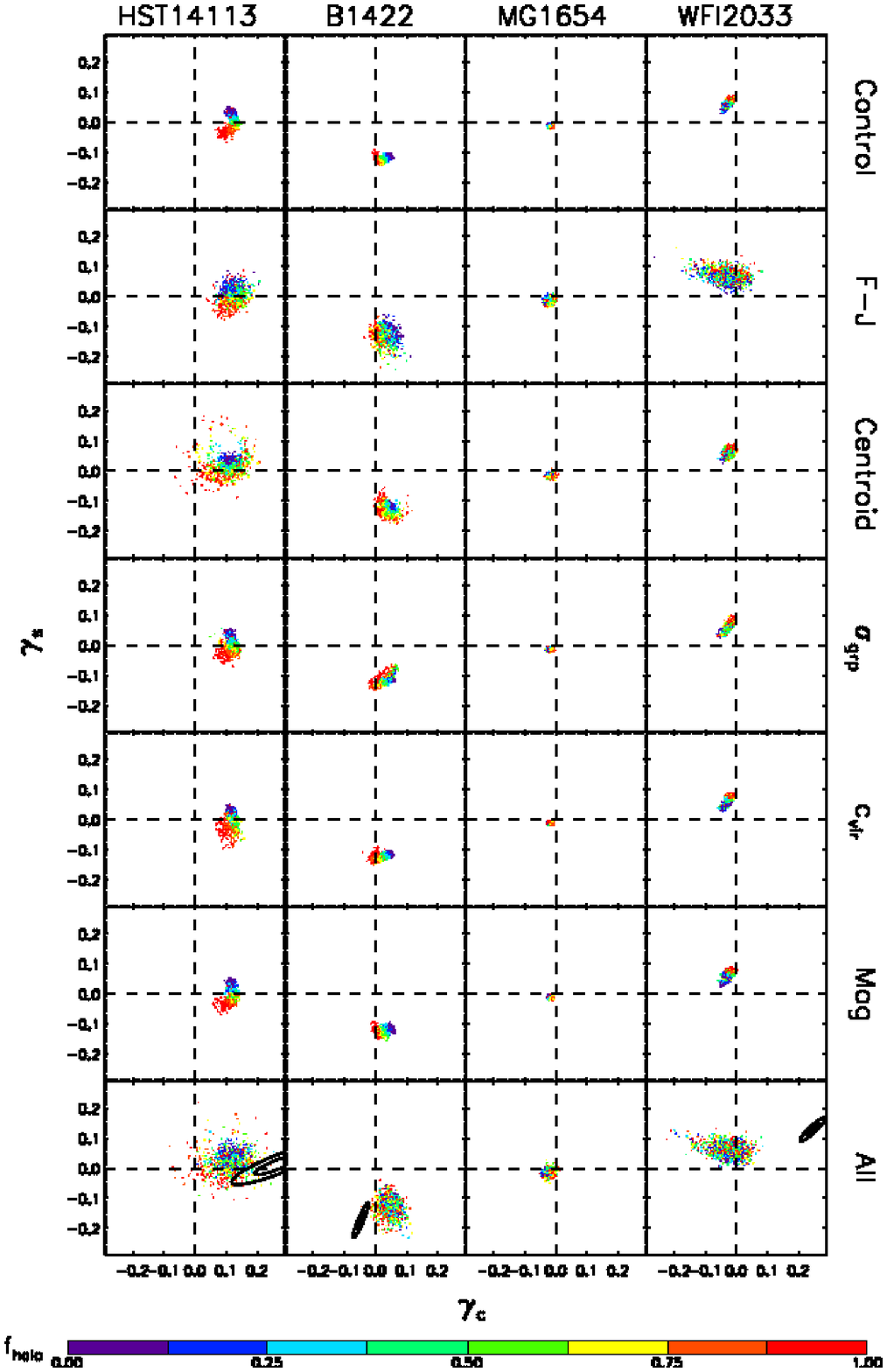}
\caption{Continued.}
\end{figure*}

The mean and scatter in $\gamma_{c}$ and $\gamma_{s}$ for each uncertainty is listed in Table~\ref{tab:err}.  We assume Gaussian errors in calculating the errors in $\gamma_{c}$ and $\gamma_{s}$, but the actual shear distributions can be asymmetric and non-Gaussian (see Figure~\ref{fig:env_calc}).  We note that:
\begin{itemize}
\item The scatter in the shear for the control sample is typically $\sim$0.01, whereas the scatter when all uncertainties are taken into account is typically $\sim$0.03.  The average increase in the scatter in either $\gamma_{c}$ or $\gamma_{s}$ is $\sim$0.015, although this varies among the individual systems.  The means of the shear distributions themselves have a typical shift of $\sim$0.01, although $\gamma_{c}$ in HE0435 changes by 0.05, the largest offset for any system.
\item The scatter in the Faber-Jackson relation and uncertainty in the group centroid position are usually the dominant sources of error.  The effect of the FJ scatter on the shear is generally more significant for low values of $f_{halo}$, when more mass is assigned to the group galaxies and their shear contribution is more greatly affected.  Conversely, the uncertainty in the group centroid position is generally more important for high values of $f_{halo}$, as the group halo has a larger effect on the shear when it is more massive.
\item The magnitude errors in the photometry are negligible, resulting in changes in the mean and standard deviation of $\gamma_{c}$ or $\gamma_{s}$ of less than 0.01.
\item The offset and increase in scatter in the shear due to the group velocity dispersion error and concentration parameter, both of which affect the group mass, are on the order of $\lesssim$ 0.01.
\item There is a large amount of scatter with little systematic dependence on $f_{halo}$.  As a result, we cannot place useful constraints on $f_{halo}$ by comparing shears from lens model results to the shears from the environment (\S~\ref{subsec:env_mod_comp}) at this time.
\item For the full lens environment, the average shear is $\langle\gamma\rangle$ = 0.08, ranging from as little as 0.02 to as much as 0.17.  The environmental effects on the lens potential are therefore comparable to the lens model shears (Table~\ref{tab:modshear}) and cannot be ignored when analyzing the lens potential.  We consider the shears due to the local group environment alone in \S~\ref{subsec:results_sep}.
\end{itemize}

\begin{table*}
\caption{Shear Statistics for Observational Uncertainties\label{tab:err}}
\begin{ruledtabular}
\begin{tabular}{l|ccccccc}
\multirow{3}{*}{Lens} &
$\langle\gamma_{c}\rangle_{control}$ &
$\langle\gamma_{c}\rangle_{FJ}$ &
$\langle\gamma_{c}\rangle_{cent}$ &
$\langle\gamma_{c}\rangle_{vd}$ &
$\langle\gamma_{c}\rangle_{c_{vir}}$ &
$\langle\gamma_{c}\rangle_{mag}$ &
$\langle\gamma_{c}\rangle_{all}$ \\ &
$\langle\gamma_{s}\rangle_{control}$ &
$\langle\gamma_{s}\rangle_{FJ}$ &
$\langle\gamma_{s}\rangle_{cent}$ &
$\langle\gamma_{s}\rangle_{vd}$ &
$\langle\gamma_{s}\rangle_{c_{vir}}$ &
$\langle\gamma_{s}\rangle_{mag}$ &
$\langle\gamma_{s}\rangle_{all}$ \\ &
$\langle\gamma\rangle_{control}$ &
$\langle\gamma\rangle_{FJ}$ &
$\langle\gamma\rangle_{cent}$ &
$\langle\gamma\rangle_{vd}$ &
$\langle\gamma\rangle_{c_{vir}}$ &
$\langle\gamma\rangle_{mag}$ &
$\langle\gamma\rangle_{all}$
\\
\tableline
\multirow{3}{*}{Q0047} &
0.06 $\pm$ 0.01 &
0.06 $\pm$ 0.01 &
0.04 $\pm$ 0.02 &
0.05 $\pm$ 0.01 &
0.06 $\pm$ 0.01 &
0.06 $\pm$ 0.01 &
0.04 $\pm$ 0.03 \\ &
0.01 $\pm$ 0.01 &
0.01 $\pm$ 0.01 &
0.01 $\pm$ 0.03 &
0.01 $\pm$ 0.01 &
0.01 $\pm$ 0.01 &
0.01 $\pm$ 0.01 &
0.01 $\pm$ 0.02 \\ &
0.06 $\pm$ 0.01 &
0.06 $\pm$ 0.01 &
0.05 $\pm$ 0.02 &
0.05 $\pm$ 0.02 &
0.06 $\pm$ 0.01 &
0.06 $\pm$ 0.01 &
0.05 $\pm$ 0.02
\\
\tableline
\multirow{3}{*}{HE0435} &
0.07 $\pm$ 0.04 &
0.07 $\pm$ 0.04 &
0.02 $\pm$ 0.04 &
0.07 $\pm$ 0.04 &
0.07 $\pm$ 0.04 &
0.07 $\pm$ 0.03 &
0.02 $\pm$ 0.04 \\ &
-0.01 $\pm$ 0.03 &
-0.01 $\pm$ 0.04 &
-0.01 $\pm$ 0.04 &
-0.01 $\pm$ 0.03 &
-0.02 $\pm$ 0.03 &
-0.02 $\pm$ 0.02 &
-0.01 $\pm$ 0.05 \\ &
0.08 $\pm$ 0.04 &
0.08 $\pm$ 0.04 &
0.05 $\pm$ 0.03 &
0.08 $\pm$ 0.04 &
0.08 $\pm$ 0.05 &
0.08 $\pm$ 0.03 &
0.05 $\pm$ 0.04
\\
\tableline
\multirow{3}{*}{MG0751} &
-0.14 $\pm$ 0.01 &
-0.15 $\pm$ 0.04 &
-0.14 $\pm$ 0.03 &
-0.14 $\pm$ 0.01 &
-0.14 $\pm$ 0.01 &
-0.14 $\pm$ 0.01 &
-0.15 $\pm$ 0.05 \\ &
-0.10 $\pm$ 0.03 &
-0.10 $\pm$ 0.03 &
-0.08 $\pm$ 0.03 &
-0.09 $\pm$ 0.02 &
-0.10 $\pm$ 0.03 &
-0.09 $\pm$ 0.03 &
-0.08 $\pm$ 0.03 \\ &
0.17 $\pm$ 0.02 &
0.18 $\pm$ 0.04 &
0.17 $\pm$ 0.03 &
0.17 $\pm$ 0.02 &
0.17 $\pm$ 0.02 &
0.17 $\pm$ 0.02 &
0.17 $\pm$ 0.05
\\
\tableline
\multirow{3}{*}{PG1115} &
-0.08 $\pm$ 0.01 &
-0.09 $\pm$ 0.01 &
-0.07 $\pm$ 0.03 &
-0.08 $\pm$ 0.01 &
-0.09 $\pm$ 0.02 &
-0.08 $\pm$ 0.01 &
-0.07 $\pm$ 0.04 \\ &
0.10 $\pm$ 0.02 &
0.10 $\pm$ 0.02 &
0.06 $\pm$ 0.03 &
0.10 $\pm$ 0.02 &
0.10 $\pm$ 0.03 &
0.10 $\pm$ 0.02 &
0.06 $\pm$ 0.03 \\ &
0.13 $\pm$ 0.02 &
0.13 $\pm$ 0.02 &
0.09 $\pm$ 0.03 &
0.13 $\pm$ 0.02 &
0.13 $\pm$ 0.03 &
0.13 $\pm$ 0.02 &
0.10 $\pm$ 0.03
\\
\tableline
\multirow{3}{*}{RXJ1131} &
-0.01 $\pm$ 0.003 &
-0.01 $\pm$ 0.01 &
-0.01 $\pm$ 0.003 &
-0.01 $\pm$ 0.003 &
-0.01 $\pm$ 0.003 &
-0.01 $\pm$ 0.003 &
-0.01 $\pm$ 0.01 \\ &
-0.01 $\pm$ 0.005 &
-0.01 $\pm$ 0.01 &
-0.01 $\pm$ 0.005 &
-0.01 $\pm$ 0.005 &
-0.01 $\pm$ 0.01 &
-0.01 $\pm$ 0.005 &
-0.01 $\pm$ 0.01 \\ &
0.01 $\pm$ 0.004 &
0.02 $\pm$ 0.01 &
0.01 $\pm$ 0.004 &
0.01 $\pm$ 0.004 &
0.01 $\pm$ 0.004 &
0.01 $\pm$ 0.004 &
0.02 $\pm$ 0.01
\\
\tableline
\multirow{3}{*}{HST14113} &
0.12 $\pm$ 0.01 &
0.12 $\pm$ 0.02 &
0.11 $\pm$ 0.04 &
0.12 $\pm$ 0.01 &
0.12 $\pm$ 0.01 &
0.12 $\pm$ 0.01 &
0.11 $\pm$ 0.04 \\ &
0.00 $\pm$ 0.02 &
0.00 $\pm$ 0.03 &
0.02 $\pm$ 0.03 &
0.00 $\pm$ 0.02 &
-0.00 $\pm$ 0.03 &
-0.00 $\pm$ 0.02 &
0.02 $\pm$ 0.04 \\ &
0.12 $\pm$ 0.01 &
0.13 $\pm$ 0.02 &
0.11 $\pm$ 0.03 &
0.12 $\pm$ 0.01 &
0.12 $\pm$ 0.01 &
0.12 $\pm$ 0.01 &
0.12 $\pm$ 0.04
\\
\tableline
\multirow{3}{*}{B1422} &
0.03 $\pm$ 0.01 &
0.03 $\pm$ 0.02 &
0.04 $\pm$ 0.02 &
0.03 $\pm$ 0.02 &
0.03 $\pm$ 0.02 &
0.03 $\pm$ 0.02 &
0.05 $\pm$ 0.02 \\ &
-0.12 $\pm$ 0.01 &
-0.13 $\pm$ 0.02 &
-0.12 $\pm$ 0.02 &
-0.12 $\pm$ 0.01 &
-0.12 $\pm$ 0.01 &
-0.12 $\pm$ 0.01 &
-0.13 $\pm$ 0.03 \\ &
0.13 $\pm$ 0.01 &
0.13 $\pm$ 0.02 &
0.13 $\pm$ 0.02 &
0.12 $\pm$ 0.01 &
0.13 $\pm$ 0.01 &
0.13 $\pm$ 0.01 &
0.14 $\pm$ 0.03
\\
\tableline
\multirow{3}{*}{MG1654} &
-0.02 $\pm$ 0.005 &
-0.02 $\pm$ 0.01 &
-0.02 $\pm$ 0.01 &
-0.02 $\pm$ 0.01 &
-0.02 $\pm$ 0.005 &
-0.02 $\pm$ 0.005 &
-0.02 $\pm$ 0.01 \\ &
-0.01 $\pm$ 0.003 &
-0.01 $\pm$ 0.01 &
-0.01 $\pm$ 0.004 &
-0.01 $\pm$ 0.004 &
-0.01 $\pm$ 0.003 &
-0.01 $\pm$ 0.003 &
-0.01 $\pm$ 0.01 \\ &
0.02 $\pm$ 0.004 &
0.02 $\pm$ 0.01 &
0.02 $\pm$ 0.01 &
0.02 $\pm$ 0.01 &
0.02 $\pm$ 0.004 &
0.02 $\pm$ 0.004 &
0.03 $\pm$ 0.01
\\
\tableline
\multirow{3}{*}{WFI2033} &
-0.02 $\pm$ 0.01 &
-0.03 $\pm$ 0.04 &
-0.02 $\pm$ 0.01 &
-0.03 $\pm$ 0.01 &
-0.02 $\pm$ 0.01 &
-0.02 $\pm$ 0.01 &
-0.03 $\pm$ 0.04 \\ &
0.06 $\pm$ 0.01 &
0.07 $\pm$ 0.02 &
0.06 $\pm$ 0.01 &
0.06 $\pm$ 0.01 &
0.06 $\pm$ 0.01 &
0.06 $\pm$ 0.01 &
0.06 $\pm$ 0.02 \\ &
0.07 $\pm$ 0.01 &
0.08 $\pm$ 0.03 &
0.07 $\pm$ 0.01 &
0.07 $\pm$ 0.01 &
0.07 $\pm$ 0.01 &
0.07 $\pm$ 0.01 &
0.08 $\pm$ 0.03
\\
\end{tabular}
\end{ruledtabular}
\tablecomments{Errors are standard deviations of the shear distributions assuming a Gaussian distribution.  In some cases, the Gaussian assumption is very rough (see Figure~\ref{fig:env_calc}), and these errors should be considered as guides only.}
\end{table*}

We also run trials where we take into account only the galaxies for which we have spectroscopic data.  Our results are qualitatively similar except in the cases of B1422 and WFI2033, two of the systems where our shears do not match the lens model results (\S~\ref{subsec:env_mod_comp}).  For these two systems, the shear distributions shift substantially but in a direction that makes them more discrepant with the lens model shears.

\subsection{Comparison of Environment and Lens Model Shears} \label{subsec:env_mod_comp}
Our shear distributions overlap the lens model-derived values in HE0435, PG1115, and HST14113, but not in RXJ1131, B1422, or WFI2033 (Figure~\ref{fig:env_calc}).  We want to quantify the level of disagreement between the lens model and environment shears, which are determined independently from one another.  To do this, we construct the probability distribution for $\Delta\gamma$, the offset between the shear distributions, to determine whether it is consistent with zero.  If we have probability distributions $p_{env}(\gamma)$ and $p_{mod}(\gamma)$ for the environment and lens model shears respectively, the formal statement of the probability distribution of $\Delta\gamma$ is 
\begin{equation}
p(\Delta\gamma) = \int p_{env}(\gamma^\prime) p_{mod}(\gamma^\prime+\Delta\gamma)d\gamma^\prime.
\end{equation}

\begin{figure*}
\centering
\plotone{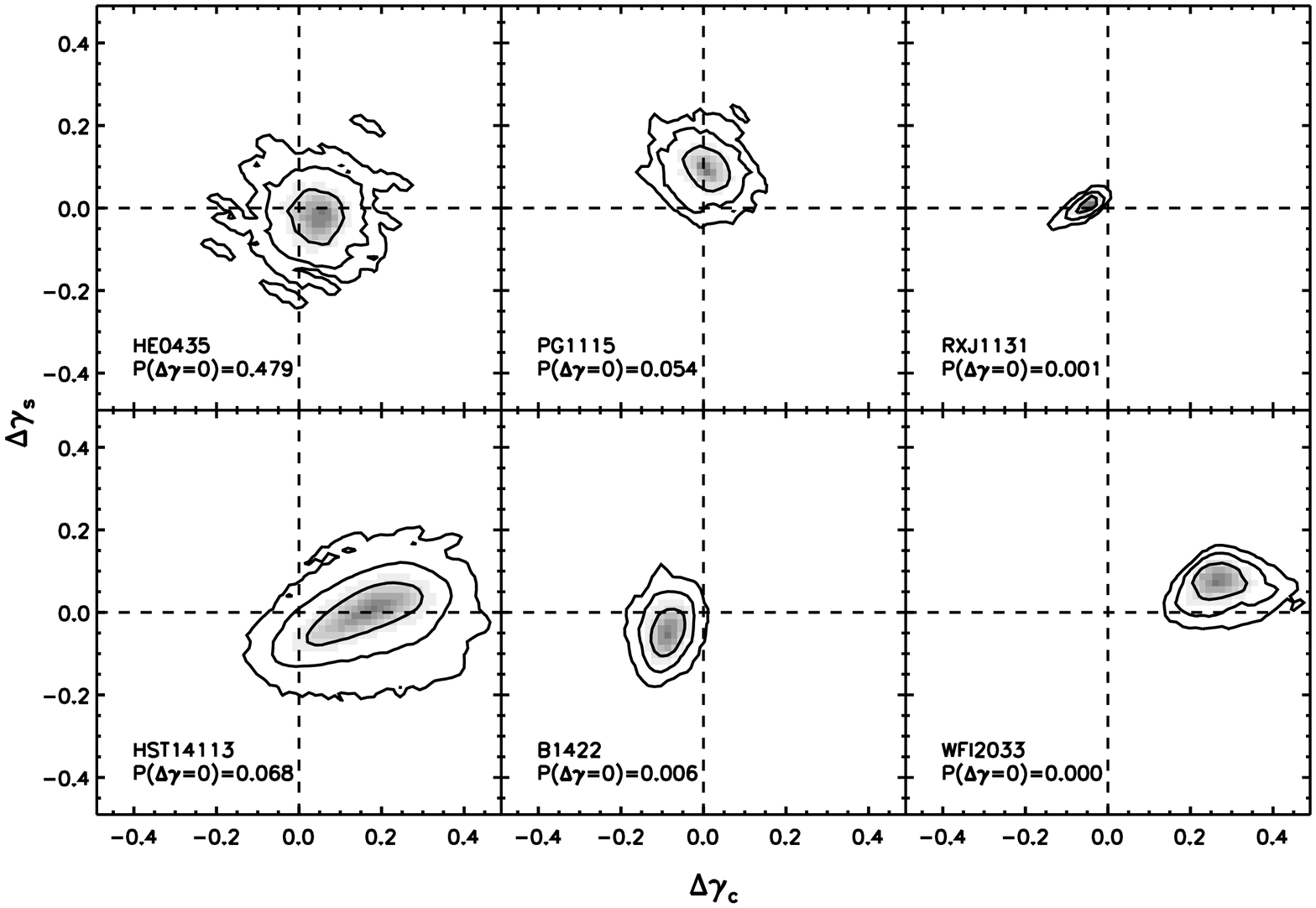}
\caption{68\%, 95\%, and 99.7\% contours for $\Delta\gamma$ in each of the six 4-image systems in our sample.  The origin represents perfect agreement between the environment shears and the lens model shears.  The $P$-value for $\Delta\gamma$ = 0 is given for each system.  A general power-law is assumed for the lens galaxy density profile in the lens modeling.  For three of the systems (HE0435, PG1115, HST14113), the environment shears are consistent with the model shears to within the 95\% confidence intervals.  However, the shears are inconsistent at greater than 95\% confidence for the remaining three systems (RXJ1131, B1422, WFI2033). \label{fig:corr_contour}}
\end{figure*}

For the environment shear, we have discrete samples rather than a continuous distribution of shears.  We can also sample 1000 discrete points in ($\gamma_{c}$,$\gamma_{s}$) space from the lens model shear distributions.  We cross-correlate these points with the 1000 shear points from the environment measurement and determine the difference between each model shear $i$ and environment shear $j$, $\Delta\gamma_{ij} = \gamma_{j}^{mod} - \gamma_{i}^{env}$.  The discrete estimator of the $\Delta\gamma$ probability distribution from these samples is then 
\begin{equation}
p(\Delta\gamma) = \frac{1}{N_{env}N_{mod}} \sum_{i=1}^{N_{env}} \sum_{j=1}^{N_{mod}} \delta(\Delta\gamma + \gamma_{i}^{env} - \gamma_{j}^{mod}).
\end{equation}
In other words, we have a collection of $N_{env}N_{mod} = 10^{6}$ points at the locations $\Delta\gamma_{ij}$ for all $i$ and $j$ in both $\gamma_{c}$ and $\gamma_{s}$ separately.  We plot these points in ($\Delta\gamma_{c}$,$\Delta\gamma_{s}$) space and calculate isoprobability contours.  The $P$-value for $\Delta\gamma=0$, which represents perfect agreement between the two distributions, is the fraction of points lying outside the contour passing through the origin.  The 68\%, 95\%, and 99.7\% contours are shown in Figure~\ref{fig:corr_contour} and the value of $P(\Delta\gamma=0$) is given for each system.

The results show that the environment and lens model shears for HE0435, PG1115, and HST14113 agree to within the 95\% contours, albeit marginally for PG1115 and HST14113.  However, the shears for RXJ1131, B1422, and WFI2033 do not.  The disagreements could arise from a number of effects that we have not considered in our environment analysis, as well as problems in the lens modeling itself.  These are discussed further in \S~\ref{subsec:biases} and \S~\ref{subsec:biases_models}.

\subsection{Constraining Line-of-Sight Shear} \label{subsec:results_sep}
\begin{figure*}
\centering
\plotone{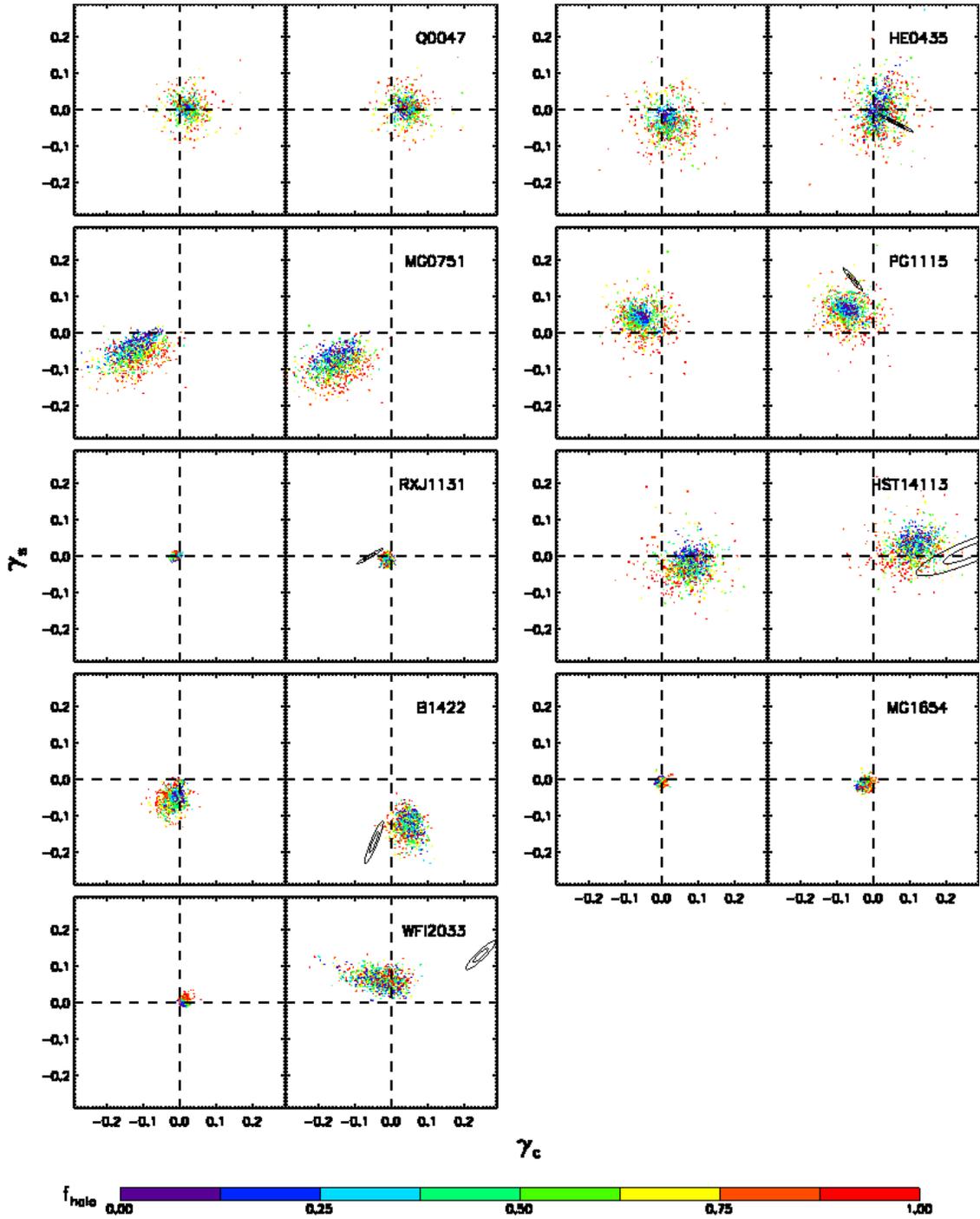}
\caption{$\gamma_{c}$ and $\gamma_{s}$ from the lens group environment alone (left panels) and from the full environment including line-of-sight galaxies (right panels).  The right panels are identical to the bottom row of Figure~\ref{fig:env_calc}.  Each point represents one of 1000 realizations of the environment, including the spectroscopic or photometrically assigned group galaxies, the LOS galaxies, and the uncertainties in Figure~\ref{fig:env_calc}.  The points are color coded by $f_{halo}$ as indicated by the color bar.  For the six 4-image lenses, the 1$\sigma$ and 2$\sigma$ error ellipses derived from lens modeling are shown for power-law lens galaxy models.  Statistics for each distribution are in Table~\ref{tab:env_calc}.  Including the LOS galaxies has a large effect on the shear in B1422 and WFI2033 due to perturbations from LOS objects projected close to the lens.  \label{fig:los_grpenv}}
\end{figure*}

We compare the shear due to the local group environment alone to that of the full environment to quantify the effects of the LOS perturbers on the shear (Figure~\ref{fig:los_grpenv}).  The local lens environment alone induces an average shear of $\gamma = 0.05$ with a range from 0.01 to 0.14, compared to that for the full environment ($\gamma = 0.08$, ranging from 0.02 to 0.17; Table~\ref{tab:env_calc}).  The shear amplitude from the LOS objects can thus be comparable to that from the local group environment, although it depends on the configuration of the individual lens systems.  We also compute $\Delta\gamma_{env}$, the mean offset in $(\gamma_{c},\gamma_{s})$ space between the shears from the local environment alone and the shears from the full environment.  The mean value of this offset across all systems is $\Delta\gamma_{env} = 0.06$ with a range from 0.02 to 0.11, indicating that the LOS can be a significant perturbation to the lens potential.  Accounting for the LOS objects does not significantly add to the shear scatter by more than 0.02 in any system.

\begin{table}
\caption{Shear Statistics From Local Group Environment and Full Environment\label{tab:env_calc}}
\begin{ruledtabular}
\begin{tabular}{l|ccc}
\multirow{3}{*}{Lens} &
$\langle\gamma_{c}\rangle_{local}$ &
$\langle\gamma_{c}\rangle_{full}$ &
\\ &
$\langle\gamma_{s}\rangle_{local}$ &
$\langle\gamma_{s}\rangle_{full}$ &
$\Delta\gamma_{env}$ \\ &
$\langle\gamma\rangle_{local}$ &
$\langle\gamma\rangle_{full}$ &
\\
\tableline
\multirow{3}{*}{Q0047} &
0.02 $\pm$ 0.03 &
0.04 $\pm$ 0.03 &
 \\ &
0.00 $\pm$ 0.02 &
0.01 $\pm$ 0.02 &
0.05 $\pm$ 0.03 \\ &
0.03 $\pm$ 0.03 &
0.05 $\pm$ 0.02 &
\\
\tableline
\multirow{3}{*}{HE0435} &
0.02 $\pm$ 0.03 &
0.02 $\pm$ 0.04 &
 \\ &
-0.03 $\pm$ 0.04 &
-0.01 $\pm$ 0.05 &
0.07 $\pm$ 0.05 \\ &
0.05 $\pm$ 0.03 &
0.05 $\pm$ 0.04 &
\\
\tableline
\multirow{3}{*}{MG0751} &
-0.12 $\pm$ 0.05 &
-0.15 $\pm$ 0.05 &
 \\ &
-0.05 $\pm$ 0.03 &
-0.08 $\pm$ 0.03 &
0.08 $\pm$ 0.04 \\ &
0.14 $\pm$ 0.05 &
0.17 $\pm$ 0.05 &
\\
\tableline
\multirow{3}{*}{PG1115} &
-0.05 $\pm$ 0.04 &
-0.07 $\pm$ 0.04 &
 \\ &
0.04 $\pm$ 0.03 &
0.06 $\pm$ 0.03 &
0.06 $\pm$ 0.04 \\ &
0.08 $\pm$ 0.03 &
0.10 $\pm$ 0.03 &
\\
\tableline
\multirow{3}{*}{RXJ1131} &
-0.01 $\pm$ 0.01 &
-0.01 $\pm$ 0.01 &
 \\ &
0.00 $\pm$ 0.005 &
-0.01 $\pm$ 0.01 &
0.02 $\pm$ 0.01 \\ &
0.01 $\pm$ 0.01 &
0.02 $\pm$ 0.01 &
\\
\tableline
\multirow{3}{*}{HST14113} &
0.07 $\pm$ 0.04 &
0.11 $\pm$ 0.04 &
 \\ &
-0.03 $\pm$ 0.04 &
0.02 $\pm$ 0.04 &
0.09 $\pm$ 0.05 \\ &
0.09 $\pm$ 0.04 &
0.12 $\pm$ 0.04 &
\\
\tableline
\multirow{3}{*}{B1422} &
-0.02 $\pm$ 0.02 &
0.05 $\pm$ 0.02 &
 \\ &
-0.05 $\pm$ 0.03 &
-0.13 $\pm$ 0.03 &
0.11 $\pm$ 0.03 \\ &
0.05 $\pm$ 0.03 &
0.14 $\pm$ 0.03 &
\\
\tableline
\multirow{3}{*}{MG1654} &
-0.00 $\pm$ 0.01 &
-0.02 $\pm$ 0.01 &
 \\ &
-0.00 $\pm$ 0.01 &
-0.01 $\pm$ 0.01 &
0.03 $\pm$ 0.01 \\ &
0.01 $\pm$ 0.01 &
0.03 $\pm$ 0.01 &
\\
\tableline
\multirow{3}{*}{WFI2033} &
0.01 $\pm$ 0.01 &
-0.03 $\pm$ 0.04 &
 \\ &
0.00 $\pm$ 0.01 &
0.06 $\pm$ 0.02 &
0.08 $\pm$ 0.03 \\ &
0.01 $\pm$ 0.01 &
0.08 $\pm$ 0.03 &
\\
\tableline
\multirow{3}{*}{Average} &
- &
- &
 \\ &
- &
- &
0.06 \\ &
0.05 &
0.08 &
 \\
\end{tabular}
\end{ruledtabular}
\tablecomments{Errors are standard deviations of the shear distributions assuming a Gaussian distribution.  In some cases, the Gaussian assumption is very rough (see Figure~\ref{fig:los_grpenv}), and these errors should be considered as guides only.}
\end{table}

\subsection{Notes on Individual Lens Systems} \label{subsec:indiv}
From the results in Figures~\ref{fig:env_calc} and~\ref{fig:los_grpenv} and quantified in Tables~\ref{tab:err} and~\ref{tab:env_calc}, it is clear that the behavior of the shear is not uniform across all lenses.  This is expected, as group environments can vary significantly in overall mass, richness, and other properties that can affect the lensing potential.  In addition, the configuration of individual galaxies relative to the lens galaxy can have large shear effects, particularly in the limit of low $f_{halo}$.  The shear from LOS galaxies can be sensitive to a small number of objects projected close to the lens, as well as to the large amount of cosmic variance in beams along lines of sight to individual systems \citep{mom06,fas10}.  In this subsection, we provide some qualitative insights about the effects of the group environments and LOS objects in individual systems, as well as previous lens model results.

The scatter in the shear distribution for RXJ1131 is relatively small.  One reason is that the lens galaxy is offset from the group centroid more than in any other group, so perturbations from the group halo and most of the other group galaxies have less of an impact on the lensing potential.  There is evidence for substructure in RXJ1131, resulting in anomalous time delays \citep{mor06,kee09,con10}.  The lens model shears become even more discrepant from our environment results if the flux ratios are not taken into account.

Accounting for LOS objects in B1422 changes the shear by $\Delta\gamma_{env} = 0.11$, the largest shift for any system.  Much of this is likely due to the presence of a bright ($I$ = 18.38), nearby LOS galaxy at a projected separation of 0.13\arcmin.

The shear from the local environment of WFI2033 alone is small, but accounting for LOS objects shifts the shear by $\Delta\gamma_{env} = 0.08$.  This is likely due to the presence of a nearby LOS object projected 0.06\arcmin~from the lens, as well as a few nearby LOS objects in our photometric catalog, similar to B1422.  There are six objects brighter than $I$ = 21.5 that are projected between 0.12\arcmin~and 0.31\arcmin~away from the lens that are not in our spectroscopic catalog.  \citet{con10} find time delay anomalies in this system, suggesting either a complex lens environment or the presence of substructure.

The scatter in the shear distributions due to uncertainty in the position of the group centroid is the dominant source of uncertainty in HE0435 and PG1115, and also has a significant effect in Q0047 and HST14113.  The panel in Figure~\ref{fig:env_calc} showing the effects of the centroid error indicate that for large values of $f_{halo}$, the shear amplitude is shifted a larger distance away from the typical shears in the corresponding control samples.  The separation between high and low values of $f_{halo}$ is not clearly defined, so we cannot constrain $f_{halo}$ based on a comparison of our shear distribution to the lens model shear distribution.  We might infer that there are more realizations of the environment with low $f_{halo}$ that match the lens model-derived shears.  Better spectroscopic sampling of the group members in this system would help to tighten the centroid position errors and reduce the scatter.  \citet{koc06} performed a modeling analysis of HE0435 assuming an ellipsoidal pseudo-Jaffe model for the lens, an SIS model for a nearby perturber (one of the objects in our photometric red sequence), plus external shear.  They find a best-fit shear amplitude of $\gamma \sim 0.05$ and $\theta_{\gamma} \approx -30^\circ$, which is consistent with our shears calculated from the environment when converted to $(\gamma_{c},\gamma_{s})$ space.  However, their results are not directly comparable to our lens model results as we do not model individual perturbers.

In the host group of HST14113, the brightest group galaxy ($I$ = 16.82) is located only 0.68\arcmin~from the lens, indicating that it could have a large effect on the shear.  However, due to the galaxy's proximity to the group centroid, its shear contribution at low values of $f_{halo}$ is approximately degenerate with the shear contribution from the group dark matter halo at large values of $f_{halo}$.  The extreme luminosity of this galaxy may be the result of blending in our photometry, but since its mass is concentrated over a narrow range of position angles relative to the lens, it is unlikely that modeling this object as the superposition of less massive galaxy halos will make a large difference in the shear.

\citet{way05} model Q0047 as an elliptical power-law model and find that the best fit is near-isothermal and requires no external shear to explain the data.  It is possible that their models could accommodate a small amount of shear, which would be consistent with our shear distribution for this system.  We find that our shear distributions are consistent with zero shear to within the 95\% confidence contours.

Most of the uncertainty in the shear in MG0751 comes from the scatter in the FJ relation, which increases the error in $\gamma$ from 0.02 to 0.04.  There is a very bright group galaxy close to the lens ($I$ = 17.89, 0.1\arcmin~separation) that is likely the main contributor to this uncertainty.  \citet{leh97} fit point mass and SIS lens models with external shear to this lens, and their shears are inconsistent with our environment shears.  However, their models do not include both ellipticity in the lens galaxy and external shear, and thus are not directly comparable to ours.

The host group of MG1654 has a velocity dispersion three times lower than most of the other groups, corresponding to a virial mass smaller by at least an order of magnitude and resulting in a small shear contribution from the group environment.  The velocity dispersion of this system may be intrinsically low, although it could be underestimated due to the small number (eight) of confirmed group galaxies in our spectroscopic sample and the lack of photometric red sequence group members.  The small shear from the group environment also leads to a very small scatter in the shear due to the observational errors we considered.  As a result, most of the shear amplitude and scatter in our environment model comes from the LOS galaxies.  A modeling analysis performed by \citet{koc95} tested a number of lens models and found that a quasi-isothermal ellipsoid ($0.9 \lesssim \alpha \lesssim 1.1)$ provided the best fit.  The range of shears and position angles obtained from his models are consistent with our shear distributions.

\subsection{Remaining Sources of Error in Environment Analysis} \label{subsec:biases}
We have modeled the environments of our lens systems, accounting for various observational and theoretical uncertainties (\S~\ref{subsec:errors}).  However, there are other sources of error that could affect our results and that may explain the discrepancy between our shear distributions and those from lens modeling (\S~\ref{subsec:env_mod_comp}), including:
\begin{itemize}
\item We assume that all galaxies in our analysis are early-type galaxies.  We apply the Faber-Jackson relation to obtain velocity dispersions and use an elliptical galaxy SED when performing bandpass corrections.  To roughly test this assumption, we estimate the fraction of red galaxies in our spectroscopic sample both by looking for emission lines in the spectra (Momcheva et al. in preparation) and comparing the observed colors to those of an Sa-type galaxy template from the GALEV2 template set \citep{bic04}.  These tests suggest that $\sim$60-80\%\footnote{This fraction is higher than the fraction of all confirmed group members that lie on the photometric red sequences (Table~\ref{tab:groups}; further details in \citet{wil06} and Williams et al. (in prepartion)), but this increase is expected given the conservative definition (i.e., much narrower color cut) of the photometric red sequence galaxies (\S~\ref{subsec:spec_incomp}).} of galaxies in the host group of the lenses and $\sim$50\% of the LOS galaxies are red, so this assumption is not baseless.  Ideally, we would like to apply the Tully-Fisher \citep[TF;][]{tul77} relation to the blue galaxies, but this is difficult because we lack information on the inclinations of disk galaxies.  In addition, the relative errors in the FJ and TF relations at intermediate redshifts are not well-constrained enough to determine a robust calibration between the two.  While we do not consider evolution in the FJ relation due to velocity dispersion evolution, we do account for passive luminosity evolution in the galaxies.

\citet{man06} find that late-type galaxies with luminosities $L \sim L_{*}$ to $2L_{*}$ have roughly half the $r$-band mass-to-light ratio of early-type galaxies at similar luminosities.  At fainter luminosities ($L \sim L_{*}/2$), the mass-to-light ratios are nearly equal.  As a crude check of the FJ assumption, we run a test where galaxies with blue colors are given half the mass (and therefore, $1/\sqrt{2}$ times the velocity dispersion) that they normally would be assigned by the FJ relation.  Blue galaxies are defined as galaxies with a $V-I$ or $R-I$ color bluer than that of the Sa-type galaxy template at the same measured or assigned redshift.  This conservative test provides intuition as to how much our shears are affected by the assumption of the FJ relation for all galaxies.  We find that this procedure changes the the shear distributions by less than 0.01 in either component and that the mismatches between our environment shears and the lens model shears remain.
\item In our analysis of the LOS, we take individual galaxy halos into account, but ignore the effects of group or cluster-sized halos.  In principle, we could perform a similar $f_{halo}$ analysis for each association along the line of sight.  We test this effect in the six fields where we have lens model shears.  We take galaxies in each galaxy group along the LOS identified by Momcheva et al. (in preparation) and replace them with an SIS halo of mass determined from the calculated group velocity dispersion.  We cross-correlate the results of these trials with those of our standard Monte Carlo trials as in \S~\ref{subsec:env_mod_comp} and find that the shears agree to within the 68\% confidence contours.  Accounting for the halos of massive LOS structures does not change the shear distributions appreciably, nor bring the shears into agreement with the lens model shears in the three cases where they are discrepant.  
\item Our assumptions in the scaling of the group galaxy halo masses could be incorrect.  Some galaxies near the centers of groups may have had their dark matter halos tidally stripped, while others on the outskirts may still have extended halos.  Using N-body simulations, \citet{ghi98} find a marginal dependence of halo extent relative to virial radius on projected separation from the center of a host galaxy cluster, suggesting that this effect is small.  \citet{lim09} perform a similar analysis and find a stronger dependence.  However, is not clear how these analyses relate to the lower mass groups in our sample.
\item We assume that the groups are all virialized, but if they are not, the velocity dispersions are imperfect measures of the group potentials.  For two of the three systems where our results do not agree with the lens model shears (B1422 and RXJ1131), we find a photometric red sequence at the group redshift, suggesting a relaxed component \citep{zab98}.
\item We assume all galaxies in our environment analysis are singular isothermal spheres for simplicity, but in reality, they are likely to be better modeled as singular isothermal ellipsoids \citep{tre06,koo06,bol08,koo09}.  This effect is probably small because the mass distribution of a perturber outside $R_{E}$ of the lens is still centered at the same projected separation and position angle from the lens.  Changes in the shape of the perturber's mass distribution will only be small effects on the potential at the lens position.  Furthermore, the fractional change in the shear can only be of order the ellipticity, and it can be an increase or decrease depending on how the perturber is oriented with respect to the lens.  It would take a fairly pathological configuration $-$ many very elongated galaxies oriented in just the right fashion $-$ to produce a dramatic change in the shear.
\item The group centroids determined from the galaxy positions could be offset from the true mass centroid.  Luminosity-weighting the centroid determination does not affect our results.  However, X-ray gas comprises up to 10\% of the total mass of the group for groups with velocity dispersions similar to the ones we consider in our sample \citep{gon07} and could be offset from the galaxy-determined centroid by $\sim10 - 100$ kpc.  The effect of this offset on the shear will be small if the projected distance from the lens to the galaxy centroid is much larger than the likely offset between any gas component and the centroid.  We run tests in B1422 and PG1115 where we fix the group halo centroid at the position of the X-ray centroid given by \citet{fas08}, but this results in the environment shears moving further from the lens model shears, indicating that this is not the cause of the discrepancy.
\item Our method for populating the LOS redshift distributions does not handle voids properly, as a given galaxy has an uniform probability of being assigned a redshift between two adjacent galaxies in a given bin, and we do not have the statistics to define voids well.  These voids could contribute a negative shear since they are underdense relative to the mean density along the line of sight.  Voids are likely less significant than overdense regions because the deviations from the mean density along the LOS are smaller than prominent peaks, which can have arbitrarily high overdensity.  Through Monte Carlo simulations, \citet{mom06} found that voids are unlikely to contribute enough negative convergence to counter that due to prominent peaks along random lines of sight.  Thus, we infer that their contribution to the shear is even less significant because cancellation effects arise from multiple voids and shear is a tensor quantity while convergence is a scalar.
\end{itemize}

\subsection{Additional Sources of Error in Lens Models} \label{subsec:biases_models}
There are several potential sources of error in the shear values derived from lens modeling, including:
\begin{itemize}
\item Modeling the lens galaxy as an ellipsoid with a single power-law density profile may not be accurate.  Two-component model fits to the density profile, encompassing both a stellar component and dark matter component, may be more realistic in certain cases \citep[e.g.][]{koc06}, as they allow for a more complex angular structure.  We fit a de Vaucouleurs + unconstrained NFW model (representing the stellar and dark matter components, respectively) to B1422 and PG1115 to test whether the shears are significantly different.  While the shear distributions shift by $\sim0.1$ compared with the power-law models, they result in PG1115 becoming inconsistent with our environment shears and do not result in agreement between the model and environment shears in B1422.
\item The lens galaxy may be tidally truncated, which would change its assumed mass distribution.  The effect of this truncation is probably small because $R_{E}$ for lens galaxies is typically $\sim$10 kpc, which is consistent with an effective radius for a massive galaxy.
\item Substructure lensing, including substructure in the lens galaxy, the group environment, and along the line of sight that is projected within $R_{E}$ of the lens galaxy, is not accounted for in the lens models.  This substructure could introduce perturbations in the lens potential \citep{mao98,met01,dal02}, resulting in astrometric perturbations in the lensed image positions \citep{che07}.  Preliminary work on HE0435 suggests that substructure does not significantly shift the model shear distributions (Fadely et al. in preparation), but all the systems need to be examined in detail to rule out substructure as a significant systematic effect in lens model shears.
\item The lens models do not account for higher order terms than $\gamma$ in the expansion of the lens potential.  Higher-order perturbations may result in better fits to the data with different values of $\gamma$ than are currently being calculated.
\item Like most previous lens modeling analyses, our lens models omit non-linear couplings between the lens plane and line-of-sight perturbers.  Including those non-linear effects might yield better fits with different values of $\gamma$.  It might also have the consequence of modifying the ``flavor'' of shear that provides the best comparison between lens models and the environment analysis (although, as noted in \S~\ref{subsec:shear_determination}, the choice of shear flavor does not significantly affect our environment-derived shear distributions).  We are currently exploring the impact of different shear flavors on lens models (Keeton et al. in preparation).
\item The lens models assume $\kappa = 0$, as standard analyses do.  This assumption is incorrect, as $\kappa > 0$ in these systems (Wong et al. in preparation).  We test the effects of this assumption by calculating the reduced shear, $g_{c,s} = \gamma_{c,s} / (1 - \kappa)$, from the $\gamma$ and $\kappa$ that we obtain from the lens environment analysis and comparing the two reduced shear components to the shears from the $\kappa \equiv 0$ lens model.  Our results do not change: the mismatches remain discrepant at the $> 95$\% level and the matches are the same as before.  As a result, and because we wish to define the environment-determined shear as the true shear, independent of the convergence, we tabulate and plot the true shears as our shears throughout this paper.
\end{itemize}

\section{CONCLUSIONS} \label{sec:conclusions}
Using new spectroscopic and photometric data from a survey of nine gravitational lenses in groups of galaxies, we quantify the effect of the local and line-of-sight environments on the lens potentials by directly constraining the environmental shear.  We also analyze the relative importance to the shear of observational and theoretical uncertainties, including uncertainty in the Faber-Jackson relation \citep{fab76}, errors in the projected group centroid position and velocity dispersion, galaxy magnitude errors in the photometric data, uncertainties in the form of the mass profile of the group halos, and uncertainty in the concentration parameter of the group halos.  The full lens environment contributes a significant shear of $\gamma$ = 0.08 on average, ranging from 0.02 to 0.17.

For the six 4-image lenses where we can compare our shears to those derived from lens modeling, our environment analysis can reproduce the lens model shears in HE0435, PG1115, and HST14113.  However, for the other three systems (RXJ1131, B1422, WFI2033), the environment shears are inconsistent with the lens model-derived values at more than the 95\% level, pointing to a serious problem in the lens modeling or in the way that we have characterized the environment.  There do not appear to be any characteristics common among the systems where our shears are inconsistent with lens models, nor do the models themselves necessarily have large $\chi^{2}$ values relative to the number of degrees of freedom.

The contribution to the shear from structures along the line of sight to the lens can have an effect on the order of that produced by the local group environment.  The average environmental shear is $\gamma$ = 0.08, compared to an average shear of $\gamma$ = 0.05 when we only consider perturbations from the local environments of the lenses.  The mean offset in $(\gamma_{c},\gamma_{s})$ space when the effects of LOS perturbers are added is $\Delta\gamma_{env}$ = 0.06.  Most of the LOS shear comes from objects projected within $\sim$2\arcmin~of the lens and brighter than $I$ = 21.5.  Less than 0.01 in either $\gamma_{c}$ or $\gamma_{s}$ is contributed on average by galaxies $\gtrsim$ 5\arcmin~away or fainter than $I$ = 21.5, so these generally can be neglected for lenses at these redshifts.  We conclude that to minimize uncertainties in the shear from spectroscopic incompleteness, future spectroscopic surveys of the environments of lenses at these redshifts should prioritize $I <$ 21.5 objects projected close (within $\sim$2\arcmin) to the lens.

Our analysis shows that accounting for many possible observational and theoretical uncertainties typically shifts the shear by $\sim$0.01 in either $\gamma_{c}$ or $\gamma_{s}$.  However, these uncertainties add $\sim$0.015 to the scatter in our shear distributions on average.  If these uncertainties are not taken into account, the scatter in the individual shear components ranges from 0.01 to 0.04 due to systematic effects alone.

Individually, scatter in the Faber-Jackson relation \citep[FJ;][]{fab76} and the error in the group centroid position contribute most to the scatter in our shear distributions, adding 0.03 to the scatter in either $\gamma_{c}$ or $\gamma_{s}$ in the most extreme cases.  Improving the errors introduced by the FJ relation requires a scaling relation with less scatter such as the Fundamental Plane \citep{djo87,ber03b,cap06,rob06}.  Obtaining effective radii for the galaxies in our sample would give better estimates of the galaxies' internal velocity dispersions.  For example, \citet{ber03b} claim an error of 0.05 in $\log{\sigma}$ and 0.01 in $\log{I_{0}}$ for the Fundamental Plane, compared to the intrinsic scatter of $\sim$0.07 in $\log{\sigma}$ that we assume for the FJ relation.  For a truncated SIS perturber, $\gamma \propto \sigma^{2}$ inside $r_{t}$ and  $\gamma \propto \sigma^{3}$ outside $r_{t}$, so a reduction in the scatter in $\sigma$ can have a large effect on the inferred shears.

The errors in the group centroid position are most significant for systems where the lens is close to the centroid, particularly HE0435, where one of the shear components changes by 0.05.  More complete spectroscopic sampling of group members and X-ray imaging of the group's hot gas distribution could reduce the centroid errors.

Magnitude errors in our photometric data contribute $< 0.01$ to the scatter in either shear component.  The errors in the group velocity dispersion propagate into errors in the total mass of the host groups, typically adding $\sim$0.01 to the shear scatter.  The errors in the concentration parameter of the group dark matter halo are similarly small.

We do not know how the mass in a group is apportioned between the group dark matter halo and the individual group galaxies because it is dependent on the degree to which the group members have been tidally stripped via galaxy-galaxy interactions as the group evolves.  Therefore, we leave $f_{halo}$, the fraction of the group's virial mass in the common group halo, as a free parameter.  Yet, even varying $f_{halo}$ does not allow us to reproduce the lens model-derived shears in half of our 4-image lens subsample.

Another theoretical uncertainty is the mass density profile of the group dark matter halo.  To test the effects of this uncertainty, we have analyzed the shear profiles of a SIS and NFW group halo.  The difference in the shear between the two mass profiles is at most $\sim$0.02 except when the lens is projected within $\sim r_{s}/2$ of the halo centroid, assuming large values of $f_{halo}$.  This difference between the mass profiles may have a small but noticeable effect for lenses with a projected separation from the group centroid near this peak, but the choice of halo profile appears to be otherwise unimportant.

The disagreement between the shears calculated directly from our environment analysis and those derived from lens modeling might arise from the fact that lens modeling generally ignores higher-order expansions of the lens potential \citep{kee04}, makes simplistic assumptions about the form of the lens galaxy mass profile, and does not account for substructure lensing \citep[e.g.][]{mao98,met01,dal02}.  There are also possible biases or sources of error that we have not considered in our environment analysis.  These effects include location-dependent tidal stripping of galaxy halos and variations in galaxy morphology.  We have now quantified the environmental perturbations on strong gravitational lens potentials further than previous studies.  Future work must characterize these other sources of uncertainties to fully understand the limits of gravitational lens constraints on important quantities like $H_{0}$ and the properties of lens galaxy halos.

Our methodology also permits us to calculate the convergence ($\kappa$) introduced by the environment (Wong et al. in preparation).  Unlike the shear, $\kappa$ is not constrained by the lens models.  While statistical constraints on $\kappa$ may be inferred from methods such as ray-tracing through cosmological simulations \citep[e.g.][]{suy10a}, measurements of the lens environments are the only way to \emph{directly} determine its effect on lens-derived quantities.  Were the convergence values for lens systems comparable to their shears (i.e. $\sim$0.1), $H_{0}$ determinations, whose errors scale as ($1 - \kappa$), would be significantly biased \citep{kee04}.  Our work shows that lensing constraints must consider both local environment and line-of-sight galaxies in their error budgets.

\acknowledgments
We thank the anonymous referee for his/her comments and suggestions, which greatly improved this paper.  We thank Yujin Yang for his input in determining the bandpass corrections used in our analysis, and Mariangela Bernardi, Daniel Christlein, Romeel Dav\'{e}, Daniel Eisenstein, Vincent Eke, Ross Fadely, Philip Hinz, Christopher Impey, and Julio Navarro for helpful discussions and input.  This work was supported by NSF grants AST-0602288 and AST-0747311.  AIZ thanks the Max-Planck-Institut für Astronomie and the Center for Cosmology and Particle Physics at New York University for their hospitality and support during her stays there.

\appendix
\section{SHEAR CALCULATION FORMALISM} \label{app:shear_formalism}
We present the details of our formalism for calculating the shear due to the host group halo (Appendix~\ref{app:sis_nfw}) and individual galaxies (Appendix~\ref{app:truncated_sis}).  The calculations are performed using an updated version of the lensing software developed by \citet{kee01}.  For the code to perform the necessary shear calculations, we must input parameters that we calculate based on our data and various assumptions that we detail in this section.

\subsection{Shear due to SIS and NFW Group Halos} \label{app:sis_nfw}
We test how large an effect the choice of an SIS or NFW profile for the group halo has on the lens potential.  The shear due to an SIS halo is 
\begin{equation} \label{eq:shear_sis}
\gamma_{SIS} = \frac{R_{E}}{2R},
\end{equation}
where R is the angular offset of the halo centroid from the lens, while $R_{E}$ is the Einstein radius given by Equation~\ref{eq:r_ein_sis}.  To calculate the shear due to an NFW halo \citep{bar96,wri00}, we need to calculate the parameters $r_{s}$ and $\kappa_{s}$.  The NFW scale radius $r_{s}$ is that at which the power law slope of the density profile is equal to that of an isothermal sphere.  $r_{s}$ is often expressed in terms of the concentration parameter, $c_{vir} = r_{vir} / r_{s}$, where $r_{vir}$ is the virial radius of the halo.  The dimensionless parameter $\kappa_{s}$ is defined as 
\begin{equation} \label{eq:kappa_s}
\kappa_{s} = \frac{r_{s} \rho_{s}}{\Sigma_{c}},
\end{equation}
where $\rho_{s}$ is the central density of the halo and $\Sigma_{c}$ is the critical surface density for lensing,
\begin{equation} \label{eq:sigma_c}
\Sigma_{c} = \frac{c^2 D_S}{4 \pi G D_{LS} D_L}.
\end{equation}
$D_L$, $D_S$, and $D_{LS}$ are the angular diameter distances between the observer and the lens, the observer and the source, and the lens and the source, respectively.  Using the radial velocities of the spectroscopically confirmed group galaxies, we estimate the velocity dispersion, $\sigma_{grp}$, and the mass of the dark matter halo, $M_{halo}$, of each group.  In the shear analysis, we add photometric red sequence galaxies to the group, and these quantities are redetermined for each Monte Carlo trial using the bootstrap method as noted in the text.  For this test, however, we assume that only the spectroscopically confirmed members are in the group.  We then calculate the parameters needed for the shear analysis.

We first approximate the virial mass of the group, $M_{vir}$ as
\begin{equation} \label{eq:m_halo}
M_{vir} = \frac{4}{3}\pi r_{vir}^{3} \Delta_c(z) \rho_{c}(z),
\end{equation}
where $\rho_{c}(z)$  and $\Delta_{c}(z)$ are the critical density and the characteristic overdensity at the lens redshift \citep{eke98} respectively.  From our assumed cosmology, we have 
\begin{equation} \label{eq:rho_crit}
\rho_{c}(z) = \frac{3 H_0^2 [\Omega_{m}(1+z)^{3}+\Omega_{\Lambda}]}{8 \pi G}.
\end{equation}
The characteristic overdensity, $\Delta_{c}$, is the ratio of the mean density inside the group halo to the critical density at the redshift of the group.  We use a form of the characteristic overdensity from \citet{eke98,eke01}:
\begin{align} \label{eq:overdensity}
\Delta_c &= 178 \left[\frac{\rho_{m}(z)}{\rho_{c}(z)}\right]^{0.45}\notag\\
&= 178 \left[\frac{\Omega_{m}(1+z)^3}{\Omega_{m}(1+z)^{3}+\Omega_{\Lambda}}\right]^{0.45}.
\end{align}
This form is accurate to within 5\% of a spherical collapse model \citep{eke98} for our cosmology.  We account for this uncertainty by allowing for a 5\% Gaussian scatter in the characteristic overdensity, which propagates directly into a 5\% error in the virial mass.  We test the effects of this uncertainty explicitly in our shear analysis and find that it is negligible in comparison with the other sources of error.

We assume the virial theorem and express the virial mass \citep[e.g.][]{kor00} as
\begin{equation} \label{eq:m_vt}
M_{vir} = \frac{3 \sigma_{grp}^2 r_{vir}}{G}.
\end{equation}
We then equate this to $M_{vir}$ from Equation~\ref{eq:m_halo} and solve for $r_{vir}$ as a function of $\sigma_{grp}$,
\begin{equation} \label{eq:r_vir}
r_{vir} = \frac{3 \sigma_{grp}}{2 \sqrt{\pi G \Delta_c \rho_c}}.
\end{equation}
Now we are able to evaluate $r_{vir}$ in terms of known or assumed quantities and can calculate $M_{halo}$ for an assumed NFW halo from Equation~\ref{eq:m_halo}.  To determine the concentration parameter $c_{vir}$ for the group halo, we use the ``ENS'' code described in \citet{eke01}.  When running this code, we assume $\sigma_{8}$ = 0.8 and a shape parameter $\Gamma = (\Omega_{m}h) e^{-\Omega_{b} - \sqrt{h / 0.5} (\Omega_{b}/\Omega_{m})} \approx$ 0.17 in our cosmology \citep{sug95}.

We now determine $r_{s}$ for each group by setting $r_{s} = r_{vir}/c_{vir}$.  To determine $\rho_s$, we set $M_{halo}$ equal to the NFW density profile integrated over a spherical volume enclosed by the virial radius.  Solving for $\rho_s$, we find
\begin{align} \label{eq:rho_s}
\rho_s &= \frac{M_{halo}}{4 \pi r_s}\left[\int_0^{r_{vir}}\frac{r}{(1+r/r_s)^2}\right]^{-1}\notag\\
&= \frac{M_{halo}}{4\pi r_{vir}^{3}} \frac{c_{vir}^{3}}{[\ln(1+c_{vir}) - c_{vir}/(1+c_{vir})]}
\end{align}
and calculate $\kappa_{s}$.  Using these parameters, we determine the shear profiles for the SIS and NFW halos for each group.  This calculation assumes that all of the group mass is in the dark matter halo, but we can rescale this by changing $M_{halo}$ to some fraction of the total virial mass as in \S~\ref{subsubsec:fhalo}.  This is not exactly correct since the concentration of the halo determined from simulations implicitly accounts for the individual galaxies as subhalos.  However, the fractional change in the concentration between trials where $M_{halo}$ = $M_{vir}$ and those where $M_{halo} << M_{vir}$ is generally smaller than our assumed scatter of $\log c_{vir}$ = 0.14 \citep{bul01,wec02}, so the rescaling of the halo mass gives reasonable results.

\begin{figure*}
\centering
\plotone{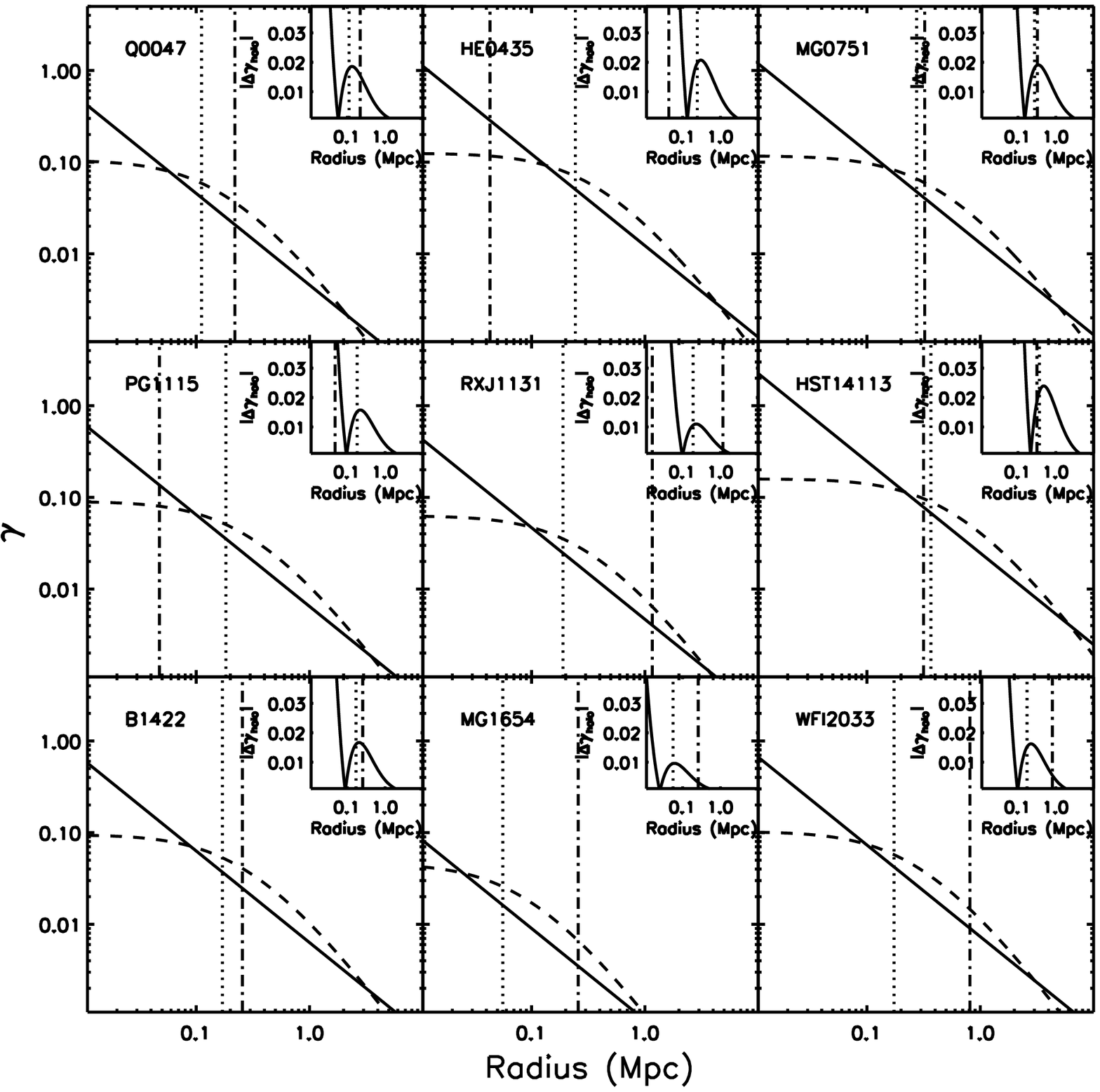}
\caption{Shear profiles for SIS (solid line) and NFW (dashed line) group halos.  The halos are assumed to contain all of the group mass.  We plot the NFW scale radius $r_{s}$ (dotted line) and the projected offset of the lens galaxy from the group centroid (dash-dotted line).  The inset panels show the absolute value of the difference in the shear profile between SIS and NFW group halos.  The peak shear difference between the two profiles is roughly $\sim$0.02 except for $r \lesssim r_{s}/2$.  Only HE0435 and PG1115 are close enough to the group centroid to fall within this region.  \label{fig:shear_profile}}
\end{figure*}

Figure~\ref{fig:shear_profile} shows the dependence of shear on projected offset from the halo center of mass for each of our nine groups.  The SIS and NFW halos are assumed to contain all the group mass, which has been calculated from the mean velocity dispersion determined across multiple Monte Carlo trials.  The absolute value of the difference between the SIS and NFW shear profiles ($|\Delta\gamma_{halo}|$ in the inset panels in Figure~\ref{fig:shear_profile}) show that the choice of profile does not make a significant difference except where the lens is projected within $\lesssim r_{s}/2$.  There is a small local peak in $|\Delta\gamma_{halo}|$ at larger radius, but its magnitude is $\sim$0.02 or less.  For low ($\lesssim$ 0.5) values of $f_{halo}$, the typical shear difference at this peak would be roughly $\sim$ 0.01 or less.  These results are consistent with the results of \citet{wri00}, who find that the mean shear ratio between NFW and SIS halos within a virial radius for a $\Lambda$CDM cosmology varies by $\approx$25\% for halo masses like ours.

\subsection{Shear due to Truncated Singular Isothermal Sphere Galaxy Halos} \label{app:truncated_sis}
In our analysis, we model the mass distribution of a galaxy-scale perturber as a truncated singular isothermal sphere (TSIS).  For a general SIS with velocity dispersion $\sigma$, the shear amplitude is given by Equation~\ref{eq:shear_sis}.  When working with TSIS profiles, we use a modified version of the basic SIS formalism.  The density profile of an isothermal sphere with velocity dispersion $\sigma$ is
\begin{equation} \label{eq:sis}
 \rho_{SIS}(r) = \frac{\sigma^2}{2\pi G r^2}.
\end{equation}
Suppose we want to truncate this profile at some radius $r_t$.  One way to obtain a smooth truncation is to write the density profile as
\begin{equation} \label{eq:tsis}
 \rho_{TSIS}(r) = \frac{\sigma^2}{2\pi G r^2} \left(1+\frac{r^2}{r_t^2}\right)^{-n}
\end{equation}
for some $n>0$.  Larger values of $n$ correspond to sharper truncation.  We can compute the lensing properties of a TSIS perturber for integer values of $n$.  For $n=6$, we find the convergence $\kappa$, shear amplitude $\gamma$, and deflection angle $\alpha$ to be:
\begin{eqnarray} \label{eq:tsis_shear}
\kappa(R) &=& \frac{R_{E}}{2R} - \frac{R_{E} D_{P}}{512 r_t} [ 256\xi^{-1} + 128\xi^{-3} + 96\xi^{-5} + 80\xi^{-7} + 70\xi^{-9} + 63\xi^{-11} ], \\
\gamma(R) &=& \frac{R_{E}}{2R} - \frac{R_{E} D_{P}}{512 r_t} [ 126(1+\xi)^{-1} + 130\xi^{-1} + 2\xi^{-3} - 30\xi^{-5} - 46\xi^{-7} - 56\xi^{-9} - 63\xi^{-11} ], \\
\alpha(R) &=& R_{E} + \frac{R_{E} r_t}{256 R D_{P}} [ -256\xi + 63 + 128\xi^{-1} + 32\xi^{-3} + 16\xi^{-5} + 10\xi^{-7} + 7\xi^{-9} ],
\end{eqnarray}
where $\xi = \sqrt{1+R^{2} D_{P}^{2}/r_t^2}$ and $D_{P}$ is the angular diameter distance to the perturber.  We assume $n=6$ throughout our analysis when discussing TSIS profiles since it represents a sufficiently sharp truncation and larger values do not make a noticeable difference in our results.

\section{TRUNCATION RADII OF GROUP GALAXIES} \label{app:trunc_rad}
When apportioning mass among the group halo and the group galaxies, the group galaxies' truncation radii are scaled so that the density at the truncation radius is the same for all group galaxies.  If we assume infinitely sharp truncation, a singular isothermal sphere with a truncation radius $r_{t}$ and velocity dispersion $\sigma$ has a density profile given by Equation~\ref{eq:sis} out to $r_{t}$.  Beyond $r_{t}$, there is no mass, so $\rho(r > r_{t}) = 0$.  The galaxy has a total mass of 
\begin{equation} \label{eq:m_sis}
M_{SIS} = \frac{2 \sigma^{2} r_{t}}{G}.
\end{equation}

Given a group with a virial mass $M_{vir}$, a halo mass fraction $f_{halo}$ (Equation~\ref{eq:f_halo}), and $N$ galaxies with masses $M_{i}$ and internal velocity dispersions $\sigma_{i}$, we want to find the truncation radii $r_{ti}$ such that $\rho(r_{ti})$ is the same for all galaxies.  We do this by setting the total mass in the group galaxies equal to the sum of the galaxy masses,
\begin{align} \label{eq:rtrunc_calc}
(1 - f_{halo}) M_{vir} &= \sum_{i=1}^{N} M_{i}\notag\\
&= \frac{2}{G} \sum_{i=1}^{N} \sigma_{i}^{2} r_{ti}\notag\\
&= \sqrt{\frac{2}{\pi G^{3} \rho_{t}}} \sum_{i=1}^{N} \sigma_{i}^{3},
\end{align}
where $\rho_{t}$ is the density at the galaxies' truncation radii.  Solving for this density, we find that
\begin{equation} \label{eq:rho_trunc}
\rho_{t} = \frac{2}{\pi G^{3} (1 - f_{halo})^{2} M_{vir}^{2}} \left[\sum_{i=1}^{N} \sigma_{i}^{3}\right]^{2}
\end{equation}
and the corresponding truncation radius of a given galaxy $j$ is 
\begin{align} \label{eq:r_trunc}
r_{tj} &= \frac{\sigma_{j}}{\sqrt{2 \pi G \rho_{t}}}\notag\\
&= \frac{1}{2} \sigma_{j} G (1-f_{halo}) M_{vir} \left[\sum_{i=1}^{N} \sigma_{i}^{3}\right]^{-1}.
\end{align}

\section{EFFECTS OF RADIAL AND LUMINOSITY SAMPLING ON LINE-OF-SIGHT SHEAR} \label{app:data_cuts}

\subsection{Projected Separation Cut} \label{app:sep_cut}
Our spectroscopic catalog for each system includes several hundred objects projected within $\sim$15\arcmin~of the lens.  However, our photometric catalog includes objects within $\sim$25\arcmin~of the lens and goes to fainter magnitudes, containing $\sim$$10^{4}$ objects per system.  With this many objects, it is not possible to achieve reasonable computation times for our Monte Carlo simulations, where we generate and analyze thousands of redshift realizations of the entire ensemble of galaxies in our catalog.  However, most of these galaxies are likely to have a negligible effect on the shear. Therefore, we cut all LOS objects with a projected separation $>$5\arcmin~from the lens, reducing the number of galaxies in our photometric catalogs by over an order of magnitude.

We expect that the shear due to LOS perturbers will decrease with projected separation from the lens for several reasons.  First, the shear from an individual perturber with an SIS profile (or any other reasonable density profile that drops off with radius) will decrease with projected separation.  Given that shear adds as a rank-2 tensor quantity, we also expect that the total shear amplitude due to all objects interior to some radius from the lens will approach some limiting value as we include more perturbers (assuming that their azimuthal positions around the lens are random).  This is the result of canceling effects from perturbers lying on lines orthogonal to each other in sky coordinates centered on the lens.  Within larger radii, we are enclosing more LOS objects, so the shear contribution from all objects near the edge of the region will tend to be smaller than objects closer to the lens where shears are less likely to cancel out.  \citet{aug07} investigated environment effects on two lenses in galaxy groups and found that their contributions were dominated by galaxies projected within 15\arcsec~of the lens, which supports our justification for a cut on projected separation.

\begin{figure*}
\centering
\plotone{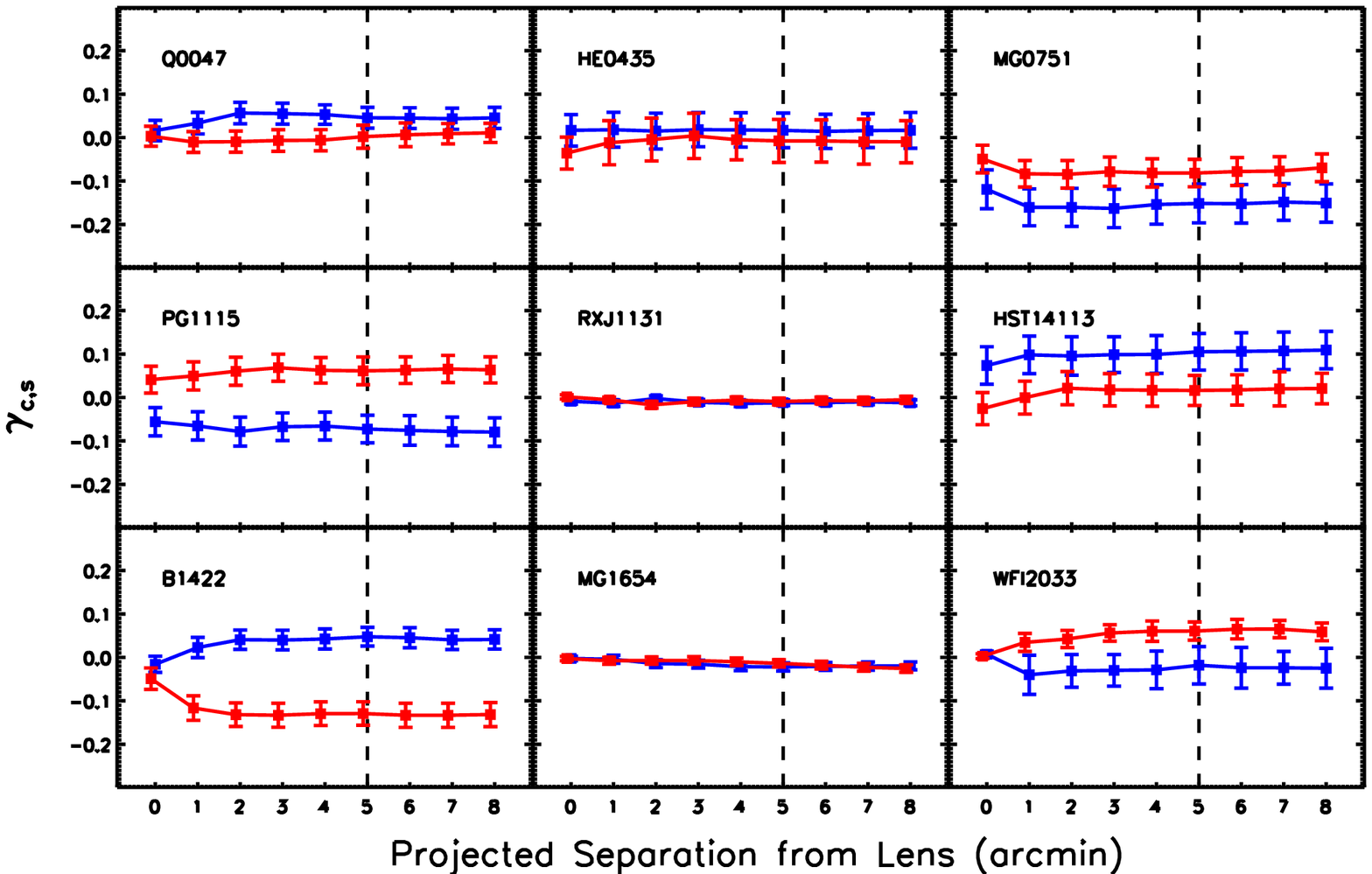}
\caption{Shear contribution from the entire local group environment plus only those LOS objects brighter than $I$ = 21.5 and within the given projected separation from the lens.  Very few objects in the local environment are fainter than $I$ = 21.5 or lie outside 5\arcmin.  The points represent the mean shear and standard deviation over 1000 trials for $\gamma_{c}$ (blue) and $\gamma_{s}$ (red).  The $\gamma_{s}$ points are offset by 0.1\arcmin~for clarity.  The first bin represents the shear from the local group environment alone.  Our projected separation cutoff is  5\arcmin~(dashed line).  Most of the variation in the shear comes from objects projected within 2\arcmin~of the lens, whereas objects projected further than 5\arcmin~away contribute less than 0.01 to the shear on average. \label{fig:los_rad_bins}}
\end{figure*}

Despite these effects, one could imagine an unlikely configuration of LOS objects arranged as to create a large shear.  To test the validity of a projected separation cut, we run our redshift randomization procedure on objects in bins of projected separation from the lens.  We perform this test for objects brighter than $I$ = 21.5, a limit that we justify in Appendix~\ref{app:mag_cut}.  For each lens, we perform 1000 trials for each of six circular apertures extending 0\arcmin~to 8\arcmin~from the lens in projected separation, each with a cutoff radius 1\arcmin~larger than the previous aperture.  In each case, we account for all objects in the local group environment, but exclude LOS objects with a projected separation outside the cutoff radius.  We then calculate the mean $\gamma_{c}$ and $\gamma_{s}$ from the objects within each aperture (Figure~\ref{fig:los_rad_bins}).  Based on the amount of variation with cutoff radius, we conclude that most of the LOS shear comes from objects projected within 2-3\arcmin~of the lens.  The mean $\gamma_{c}$ and $\gamma_{s}$ contribution from the objects beyond 5\arcmin~across all nine lens environments is smaller than 0.01, well within the error bars.  This is consistent with \citet{bra10}, who finds that the shear due to galaxy-galaxy weak lensing extrapolates to zero at a separation of $\sim$5\arcmin.  Based on this analysis, we assume that 5\arcmin~is a reasonable, conservative cut-off in projected separation.

\subsection{Magnitude Cut} \label{app:mag_cut}
For LOS objects in our photometric sample that are fainter than our spectroscopic limit of $I$ = 21.5, our method for creating a redshift distribution based on objects in our spectroscopic catalog (\S~\ref{subsec:spec_incomp}) breaks down because we do not have a representative spectroscopic sample from which to draw redshifts.  Objects in our photometric catalog can be as faint as $I \sim$ 24, but the star-galaxy separation routine described in \citet{wil06} starts to break down around $I \sim$ 21.5.  One solution, which we employ, is to cut objects fainter than $I$ = 21.5 from our LOS sample.  The main assumption is that fainter objects contribute little to the shear in comparison to brighter objects.  Objects fainter than this limit are either less massive than the typical galaxy at the low-mass end of our spectroscopic sample, or at higher redshift than the lens.  Both effects will reduce their shear contribution.

\begin{figure*}
\centering
\plotone{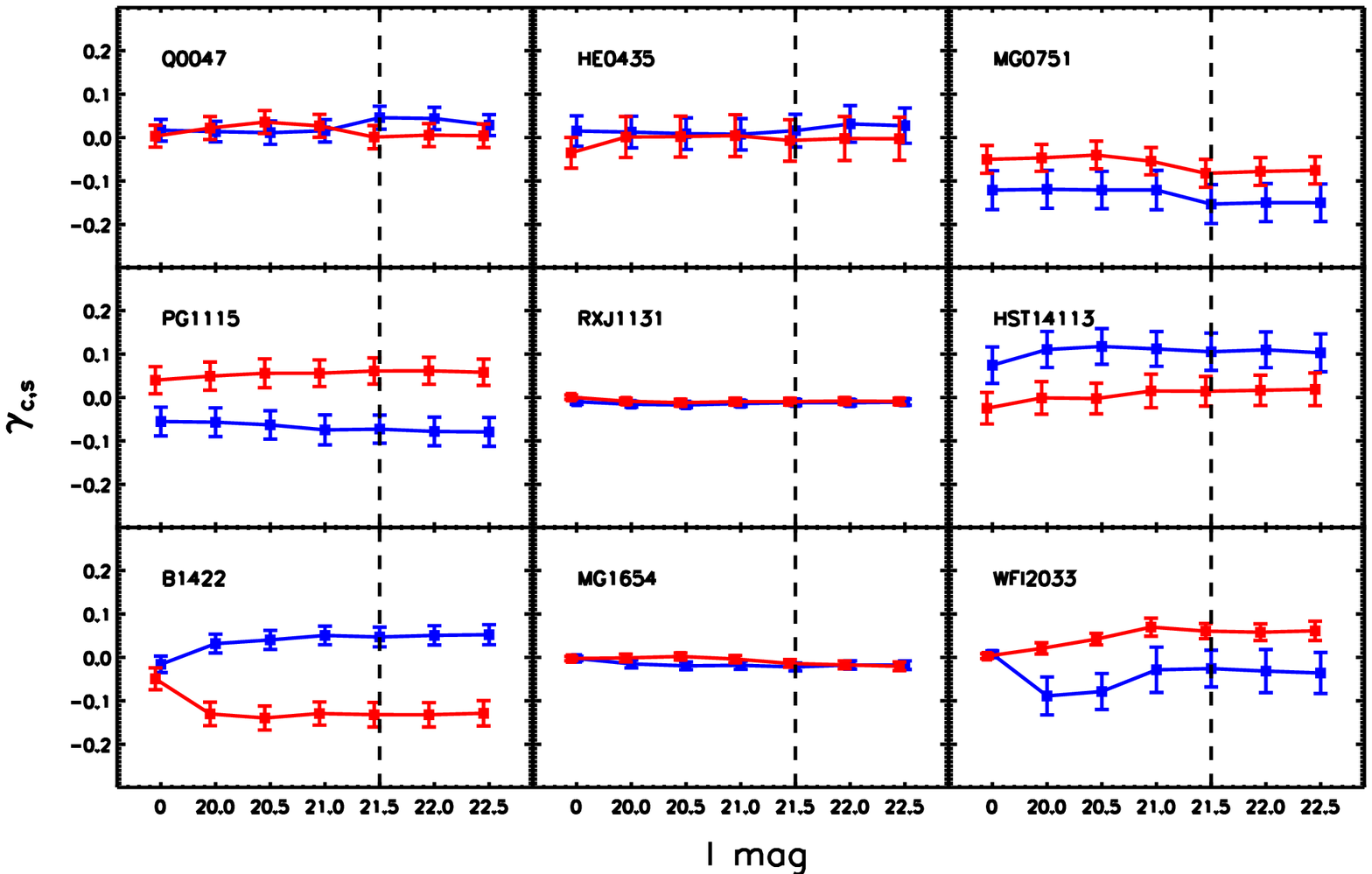}
\caption{Shear contribution from the enitre local group environment plus only those LOS objects projected within 5\arcmin~of the lens and brighter than the given $I$ magnitude.  Very few objects in the local environment are fainter than $I$ = 21.5 or lie outside 5\arcmin.  The points represent the mean shear and standard deviation over 1000 trials for $\gamma_{c}$ (blue) and $\gamma_{s}$ (red).  The $\gamma_{s}$ points are offset by 0.1 mags for clarity.  The first bin represents the shear from the local group environment alone.  Our magnitude cutoff is $I$ = 21.5 (dashed line).  Most of the variation in the shear due to the LOS comes from objects brighter than 21.5, while fainter objects contribute less than 0.01 to the shear on average.  \label{fig:los_mag_bins}}
\end{figure*}

To justify the $I$ = 21.5 cut, we plot the shears from the local group environment plus all LOS galaxies within 5\arcmin~of each lens as a function of cutoff magnitude (Figure~\ref{fig:los_mag_bins}).  In each bin, we only include the shear from LOS galaxies brighter than some limiting magnitude all the way down to $I$ = 22.5.  For the bins fainter than $I$ = 21.5, we draw redshifts from our faintest galaxy redshift distributions (roughly $I$ = 21 - 21.5) in \S~\ref{subsec:spec_incomp}.  We run 1000 trials for each bin, plotting the mean and standard deviations of $\gamma_{c}$ and $\gamma_{s}$.  Because our redshift assignment procedure draws redshifts based on a distribution of galaxies limited at $I$ = 21.5, it will tend to underestimate the masses of galaxies fainter than this limit.  However, this effect will be offset to some extent by the fact that their true redshift distribution is likely shifted away from the lens redshift to higher redshifts, reducing their shear contribution.

The results of Figure~\ref{fig:los_mag_bins} show that much of the LOS shear comes from bins brighter than $I$ = 21.5 (our spectroscopic limit).  At fainter magnitudes, the difference in the shear components is $\sim$0.01.  As with the projected separation, there tends to be a decline in additional shear contributions as we go to fainter magnitudes, and the variations are well within our error bars beyond $I$ = 21.5.  For this analysis, we adopt this conservative magnitude cut to be consistent with the limit of our spectroscopic catalog.

%\bibliography{ms}

\end{document}